\documentclass[useAMS,usenatbib,usegraphicx]{mn2e}
\title[Be~27, Be~34, and Be~36]{The anticentre old open clusters Berkeley 27, Berkeley 34, and Berkeley 36: new additions to the BOCCE project.\thanks{ This work is based on data collected at ESO telescopes under programme 076.D-0119.}}
\author[P. Donati, A. Bragaglia, M. Cignoni, G. Cocozza, and M. Tosi]{P. Donati$^{1,2}$, A. Bragaglia$^{2}$, M. Cignoni$^{1,2}$, G. Cocozza$^{2}$, and M. Tosi$^{2}$
\\
$^{1}$Dipartimento di Astronomia, via Ranzani 1, 40127 Bologna, Italia\\
$^{2}$INAF-Osservatorio Astronomico di Bologna, via Ranzani 1, 40127 Bologna, Italia\\
}
\begin{document}


\pagerange{\pageref{firstpage}--\pageref{lastpage}} 

\maketitle

\label{firstpage}

\begin{abstract}
In this paper we present the investigation of the evolutionary status of three open clusters: Berkeley 27, Berkeley 34, and Berkeley 36, all located in the Galactic anti-centre direction. All of them were observed with SUSI2@NTT using the Bessel B, V, and I filters. The cluster parameters have been obtained using the synthetic colour-magnitude diagram (CMD) method \textit{i.e.} the direct comparison of the observational CMDs with a library of synthetic CMDs generated with different evolutionary sets (Padova, FRANEC, and FST). This analysis shows that Berkeley 27 has an age between 1.5 and 1.7 Gyr, a reddening $E(B-V)$ in the range 0.40 and 0.50, and a  distance modulus $(m-M)_0$ between 13.1 and 13.3; Berkeley 34 is older with an age in the range 2.1 and 2.5 Gyr, $E(B-V)$ between 0.57 and 0.64, and $(m-M)_0$ between 14.1 and 14.3; Berkeley 36, with an age between 7.0 and 7.5 Gyr, has a reddening $E(B-V)\sim0.50$ and a distance modulus $(m-M)_0$ between 13.1 and 13.2. For all the clusters our analysis suggests a sub-solar metallicity in accord with their position in the outer Galactic disc.    
\end{abstract}


\section{Introduction}
\label{sec:intro}
This paper is part of the BOCCE (Bologna Open Clusters Chemical Evolution) project, described in detail by \cite{boc_06}. The aim of the project is to precisely and homogeneously derive the fundamental properties of a large, significant
sample of Open Clusters (OCs). OCs are among the best tracers of the properties of the Galaxy (e.g. \citealt{fri_95}). They can be used to get insight on the formation and evolution of the Galactic disc(s), the final goal of the BOCCE project. We have already published results based on photometry for 26 OCs (see \citealt{boc_06,cig_11},  and references therein), concentrating on the old ones, the most important to study the early epochs of the Galactic discs.

The three clusters examined in this paper are Berkeley 27 (also known as Biurakan~11 and hereafter Be~27 with Galactic coordinates $l=207.8^{\circ},b=2.6^{\circ}$), Berkeley 34 (also known as Biurakan~13 and hereafter Be~34, $l=214.2^{\circ},b=1.9^{\circ}$), and Berkeley 36 (Be~36, $l=227.5^{\circ},b=-0.6^{\circ}$). They are all located in the anti-centre direction, very close to the Galactic plane and have an age older than 1 Gyr.  They were selected because, based on literature studies, they all lie beyond a Galactocentric distance of 10 kpc, hence they can be useful to understand the properties of the outer disc. In particular they are located in the region where the radial metallicity distribution changes its slope and where more clusters should be studied to better understand why this happens (see, e.g.,  \citealt{sest_08,fri_10,and_11,lep_11}).  
These OCs have already been studied to different degrees in the past: the resulting parameters sometimes agree with each other and sometimes not. We present here their $BVI$ photometry, used to improve upon previous determinations of their parameters using the CMD synthetic method, as done throughout the BOCCE series.

All three clusters have been studied by \cite{hase_04} as part of a survey of 14 anti-centre clusters; they obtained $BVI$ photometry with a 0.65-m telescope. Be~27 has also been studied by \cite{car_07} using $VI$ photometry acquired at a 0.9-m telescope. Be~34 and Be~36 have been observed also by \cite{orto_05} at a 3.5-m telescope using the $BV$ filters. 
In all the three papers, the clusters parameters have been derived using isochrone fitting.

Concerning Be~27, \cite{hase_04} find a cluster age of 2.0 Gyr, a mean Galactic reddening $E(V-I)=0.30$ (or $E(B-V)=0.24$), a distance modulus of $(m-M)_0=14.25$, and a metallicity of $Z=0.03$; however, according to them, some ambiguity in the photometric calibration could have hampered the interpretation of the data. \cite{car_07} confirm a cluster age of 2.0 Gyr, but prefer a higher reddening of $E(B-V)=0.35$ and a distance modulus $(m-M)_0=14.30$; they used the Padova tracks with solar metal abundance  ($Z=0.019$). The cluster lacks a clear red giant branch and clump, which makes  the analysis of the cluster more uncertain.

For Be~34, \cite{hase_04} find a cluster age of 2.8 Gyr, a mean reddening $E(V-I)=0.60$ (i.e., $E(B-V)=0.48$), a distance modulus  $(m-M)_0=15.80$, and a metallicity $Z=0.019$. \cite{orto_05} suggest two different interpretations with two different metallicities: 2.3 Gyr, $E(B-V)=0.30$, and $(m-M)_0=15.4$ for Padova isochrones with $Z=0.019$; 2.3 Gyr, $E(B-V)=0.41$ and $(m-M)_0=15.62$ for  $Z=0.008$. Be~34 has not a clear clump either and the contamination of field stars is important, conditions that put more uncertainties on the cluster parameters estimation.

In the case of Be~36, \cite{hase_04} find a cluster age of 3.4 Gyr, a reddening of $E(V-I)=0.55$ (i.e., $E(B-V)=0.44$), a distance modulus of $(m-M)_0=15.30$ and a metallicity $Z=0.019$. They could not firmly define the clump as the cluster shows a blurred and heavily contaminated CMD, therefore they adopted the solution that could fit appropriately the main sequence and the red giant branch. \cite{orto_05} present two different cluster parameter estimations using the Padova tracks with $Z=0.019$ and with $Z=0.008$. For the solar metallicity they find a cluster age of 4 Gyr, $E(B-V)=0.25$, and $(m-M)_0=14.70$; for the sub-solar metallicity they find $E(B-V)=0.36$, $(m-M)_0=14.85$, and an age of 4 Gyr. They chose different main sequence turn-off and red clump levels with respect to \cite{hase_04}, and this can explain the differences in the results obtained.

This paper is organised as follows. Observations and the resulting CMDs are presented in Section~\ref{sec:data}; the estimation of the clusters centre in Section~\ref{sec:centre}; the derivation of their age, distance, reddening, and metallicity using comparison to synthetic CMDs in Section~\ref{sec:CMDsynth}. Discussion and summary can be found in Section~\ref{sec:sum}.

\section[]{The Data}
\label{sec:data}

\subsection{Observations}
\begin{figure}
\includegraphics[width=1.0\linewidth]{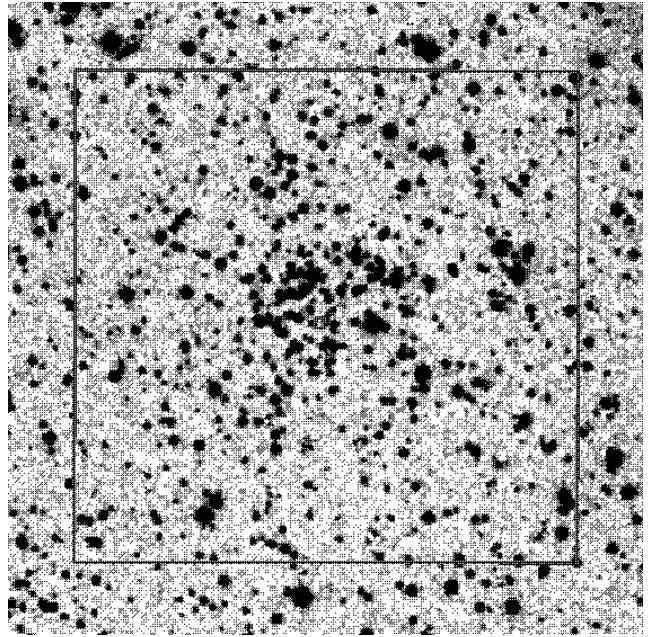}
\caption{DSS image of the field of view centred on Be~27. The box is the composite FoV of SUSI2 obtained with the rotator in different positions: only the stars inside the smaller central box fell in the mosaic gap.}
\label{fig:fovbiu11}
\end{figure}

\begin{figure}
\includegraphics[width=1.0\linewidth]{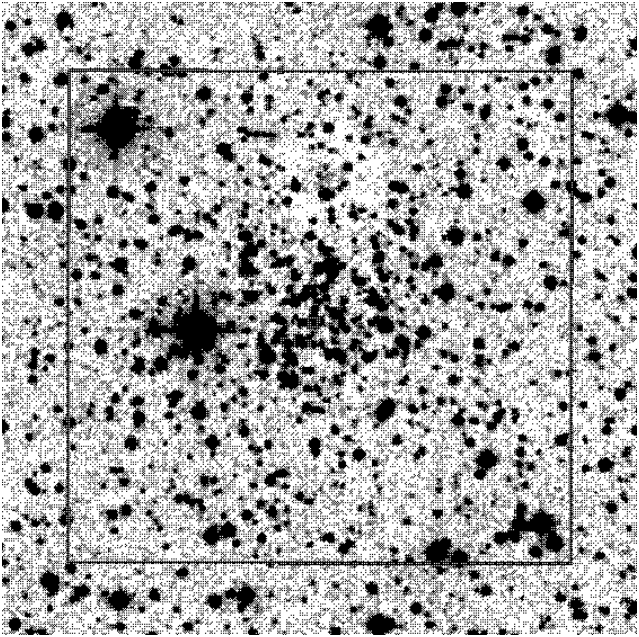}
\caption{Same as Fig.~\ref{fig:fovbiu11} but for Be~34.}
\label{fig:fovbiu13}
\end{figure}

\begin{figure}
\includegraphics[width=1.0\linewidth]{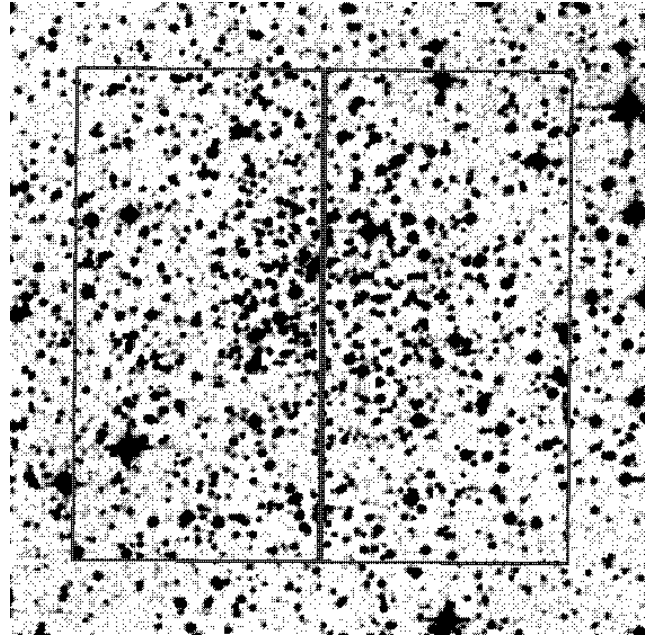}
\caption{Same as Fig.~\ref{fig:fovbiu11} but for Be~36. In this case only the default orientation was
used and the FoV has a gap apparent in the figure.}
\label{fig:fovbe36}
\end{figure}

The three clusters were observed in service mode at the ESO 3.58-metre New Technology Telescope (NTT) of the La Silla Observatory (Chile) with the instrument SUperb Seeing Imager (SUSI2) in 2005 and 2006. The instrument was composed by a mosaic of two EEV CCDs (2048$\times$4096 pixels) placed in a row. The field of view (FoV) of SUSI2 is equivalent to 5.5$\times$5.5 arcmin$^2$, with a pixel scale of 0.085 arcsec pixel$^{-1}$; for these observations the instrument was set in the 2$\times$2 binned mode (pixel scale 0.161 arcsec). The data were collected with the B, V, and I Bessel filters. The clusters were positioned at the geometric centre of the mosaic with the rotator in the default position; two of them were also observed with the instrument rotated 90 deg clockwise in order to recover stars falling in the mosaic gap. Digitized Sky Survey (DSS) images of the SUSI2 FoV for the pointings of Be~27, Be~34, and Be~36 are shown in Figures~\ref{fig:fovbiu11},~\ref{fig:fovbiu13},~and~\ref{fig:fovbe36}. The observations log-book for the three clusters is presented in Table~\ref{tab:log}. Comparison fields were also observed for decontamination purposes, located 30 arcmin away from the cluster centre (see Table~\ref{tab:log}). The seeing was below 1.5\arcsec \ for all images and below 1\arcsec \ for many. For each cluster observations in photometric condition were obtained which allowed a proper calibration using the photometric standard fields SA98, SA101-262, PG0918, and RU152 \citep{lan_92}.

\begin{table*}
 \centering
 \begin{minipage}{140mm}
  \caption{Log of observations.}
  \label{tab:log}
  \begin{tabular}{@{}lccccccc@{}}
\hline
\hline
Cluster & RA & Dec & Date & Rot\footnote{Angle of the rotator: ``a'' means default position, ``b'' means 90 deg clockwise.} & B & V & I \\
 & (J2000) & (J2000) & & & exptime & exptime & exptime \\
\hline
Be~27 & 6 51 21 & +05 46 00 & Nov, 29 2005 & a & 10s, 44s, 540s & 5s, 270s & 5s, 270s \\
 & & & & b & 10s, 540s & 5s, 270s & 5s, 270s \\
 & & & Feb, 19 2006 & a  & 50s, 100s & 50s, 50s & 50s, 50s \\
 & & & & b & 100s & 50s & 50s \\
\hline
Be~27 ext &  6 51 21 & +05 18 00 & Nov, 29 2005 & a & 2$\times$10s, 560s & 5s, 280s & 5s, 280s \\
 & & & Feb, 19 2006 & a & 100s & 50s & 50s \\
\hline
Be~34 & 7 00 23 & -00 14 11 & Nov, 29 2005 & b & 10s, 103s, 540s & 5s, 270s & 5s, 270s \\
 & & & Jan, 26 2006 & a & 10s, 540s & 2$\times$5s, 270s & 2$\times$5s, 270s \\
 & & & & b & 4$\times$10s, 2$\times$540s & 4$\times$5s, 2$\times$270s & 4$\times$5s, 2$\times$270s \\
 & & & Feb, 24 2006 & a & 100s & 50s & 50s \\
 & & & & b & 100s & 50s & 50s \\
\hline
Be~34 ext & 7 00 23 & -00 50 11 & Nov, 29 2005 & a & 10s, 77s, 560s & 5s, 280s & 5s, 280s \\
 & & & Feb, 24 2006 & a & 100s & 50s & 50s \\
\hline
Be~36 & 7 16 24 & -13 11 50 & Jan, 26 2006 & a & 10s, 540s & 5s, 270s & 2$\times$5s, 270s \\ 
 & & & Feb, 25 2006 & a & 100s & 50s & 50s \\
 & & & Mar, 20 2006 & a & 100s & 50s & 50s \\
\hline
Be~36 ext & 7 16 24 & -13 42 00 & Jan, 26 2006 & a & 10s, 560s & 5s, 280s & 5s, 280s \\
 & & & Feb, 25 2006 & a & 100s & 50s & 50s \\
 & & & Mar, 20 2006 & a & 100s & 50s & 50s \\
\hline
\end{tabular}
\end{minipage}
\end{table*}

\subsection{Data reduction}
Bias and flat field corrections were done using a standard analysis with IRAF\footnote{IRAF is the Image Reduction and Analysis Facility, a general purpose software system for the reduction and analysis of astronomical data. IRAF is written and supported by the IRAF programming group at the National Optical Astronomy Observatories (NOAO) in Tucson, Arizona.}.

The source detection and relative photometry were performed independently on each B, V, and I image, using the PSF-fitting code DAOPHOTII/ALLSTAR \citep{ste_87,ste_94}. For each frame a sample (20 to 70) of isolated and bright stars was selected to compute the PSF. The profile-fitting algorithm was imposed to determine a spatially variable PSF to include a quadratic dependence on the $x$ and $y$ coordinates in order to minimise geometrical distortion biases. Different exposure times let us recover efficiently bright and faint stars.

The next step was to remove any systematic difference between the magnitude scale of all the frames. The stars in each frame were first matched to the ones taken in photometric conditions using DAOMATCH and DAOMASTER. Then the average and the standard error of the mean of the independent measures obtained from the different images were adopted as the final values of the instrumental magnitude and uncertainty.

About 20 standard areas ($\sim6$ per filter) were observed during each photometric night and the magnitude of the standard stars was measured. Three sets of calibration equations were derived, as the targets were observed in different nights. The results are reported in Table~\ref{tab:eqcal}. 

\begin{table}
 \centering
  \caption{Calibration equations. $B$, $V$, and $I$ are the magnitudes in the standard Johnson-Cousins system while $b$, $v$, and $i$ are the instrumental magnitudes.}
  \label{tab:eqcal}
  \begin{tabular}{@{}lccc@{}}
\hline
\hline
Cluster & Date & Equation & r.m.s. \\
\hline
Be~27 & Feb, 19 2006 & $B=b-0.143(b-v)+0.275$ & 0.017 \\ 
 & & $V=v-0.016(b-v)+0.562$ & 0.024 \\
 & & $V=v-0.019(v-i)+0.540$ & 0.023 \\
 & & $I=i-0.016(v-i)-0.450$ & 0.019 \\
\hline
Be~34 & Feb, 24 2006 & $B=b-0.125(b-v)+0.312$ & 0.023 \\
 & & $V=v-0.010(b-v)+0.620$ & 0.023 \\ 
 & & $V=v-0.014(v-i)+0.602$ & 0.019 \\
 & & $I=i-0.035(v-i)-0.376$ & 0.022 \\
\hline
Be~36 & Feb, 25 2006 & $B=b-0.129(b-v)+0.318$ & 0.015 \\
 & & $V=v-0.016(b-v)+0.631$ & 0.019 \\
 & & $V=v-0.015(v-i)+0.606$ & 0.018 \\
 & & $I=i-0.024(v-i)-0.367$ & 0.018 \\
\hline
\end{tabular}
\end{table}

As the photometry of the standard stars was computed using aperture photometry, the instrumental magnitudes of the scientific targets were corrected to match the standard Johnson-Cousins system and then calibrated.
Two different calibration equations were derived for the $V$ magnitude: one using the $(B-V)$ colour index and the other one using the $(V-I)$ colour index. The difference between the two calibrations is, on average, well below one hundredth of magnitude with a small dispersion and a very shallow dependence on colour.

The \textit{Guide Star Catalogue 2.3} was used to find an accurate astrometric solution to transform the instrumental pixels positions into J2000 celestial coordinates. More than 200 stars were used for each frame as astrometric standards and the final transformations, obtained with the code CataXcorr\footnote{CataXcorr was developed by Paolo Montegriffo at INAF - Osservatorio Astronomico di Bologna.}, has an r.m.s. scatter less than 0.2\arcsec in both RA and Dec.

The final step of the data reduction process consisted in recovering the completeness level of the photometry. The procedure is the classical one  consisting of an extensive artificial stars experiment, already used in our previous works
\cite[see e.g.][for a description]{bel_02}. About 50000 stars have been artificially added and uniformly distributed on the deepest frames in groups of about 120 stars at a time, to avoid changing the actual crowding conditions. For each iteration of the artificial stars experiment the frames were reduced using the same reduction process described above. The fraction of recovered stars at different magnitude levels represents the completeness of our photometry; values are presented in Table~\ref{tab:compl}.

\begin{table*}
 \centering
 \begin{minipage}{140mm}
  \caption{Completeness of the photometry for the three clusters.}
  \label{tab:compl}
  \begin{tabular}{@{}cccccccccccc@{}}
\hline
\hline
 mag & compl $B$ & compl $V$ & compl $I$ & & compl $B$ & compl $V$ & compl $I$ & & compl $B$ & compl $V$ & compl $I$ \\
 & \multicolumn{3}{c}{Be~27} & & \multicolumn{3}{c}{Be~34} & & \multicolumn{3}{c}{Be~36} \\
\hline
16.0 & 100$\pm$6 & 99$\pm$5 & 100$\pm$4 & & 100$\pm$6 & 100$\pm$6 & 99$\pm$5 & & 100$\pm$6 & 99$\pm$5 & 99$\pm$4 \\
16.5 &  99$\pm$6 & 98$\pm$3 & 100$\pm$3 & & 100$\pm$5 & 100$\pm$5 & 99$\pm$4 & & 100$\pm$5 & 98$\pm$5 & 98$\pm$4 \\
17.0 &  99$\pm$4 & 99$\pm$3 &  99$\pm$3 & & 100$\pm$5 &  98$\pm$4 & 98$\pm$3 & &  99$\pm$5 & 99$\pm$4 & 98$\pm$3 \\
17.5 &  99$\pm$3 & 98$\pm$2 &  98$\pm$3 & &  98$\pm$4 &  98$\pm$4 & 97$\pm$3 & &  99$\pm$5 & 99$\pm$4 & 98$\pm$2 \\
18.0 &  98$\pm$3 & 99$\pm$2 &  98$\pm$2 & &  99$\pm$4 &  98$\pm$3 & 97$\pm$2 & & 100$\pm$4 & 98$\pm$3 & 97$\pm$2 \\ 
18.5 &  98$\pm$2 & 97$\pm$2 &  96$\pm$2 & &  98$\pm$3 &  98$\pm$3 & 96$\pm$2 & &  98$\pm$4 & 97$\pm$2 & 98$\pm$2 \\
19.0 &  98$\pm$2 & 97$\pm$2 &  96$\pm$2 & &  99$\pm$3 &  97$\pm$2 & 94$\pm$2 & &  97$\pm$3 & 98$\pm$2 & 96$\pm$2 \\
19.5 &  97$\pm$2 & 96$\pm$2 &  96$\pm$2 & &  98$\pm$2 &  97$\pm$2 & 94$\pm$2 & &  98$\pm$2 & 97$\pm$2 & 95$\pm$2 \\
20.0 &  98$\pm$2 & 96$\pm$2 &  93$\pm$2 & &  97$\pm$2 &  96$\pm$2 & 90$\pm$2 & &  98$\pm$2 & 96$\pm$2 & 94$\pm$2 \\
20.5 &  96$\pm$2 & 95$\pm$2 &  88$\pm$2 & &  96$\pm$2 &  95$\pm$2 & 74$\pm$2 & &  96$\pm$2 & 95$\pm$2 & 88$\pm$2 \\
21.0 &  95$\pm$2 & 93$\pm$2 &  80$\pm$2 & &  95$\pm$2 &  92$\pm$2 & 38$\pm$3 & &  96$\pm$2 & 94$\pm$2 & 75$\pm$2 \\
21.5 &  94$\pm$2 & 88$\pm$2 &  59$\pm$2 & &  92$\pm$2 &  87$\pm$2 &  5$\pm$7 & &  94$\pm$2 & 89$\pm$2 & 42$\pm$3 \\
22.0 &  88$\pm$2 & 78$\pm$2 &  36$\pm$3 & &  72$\pm$2 &  50$\pm$3 &  0       & &  84$\pm$2 & 71$\pm$2 &  7$\pm$6 \\
22.5 &  74$\pm$2 & 59$\pm$2 &  13$\pm$5 & &  17$\pm$4 &   8$\pm$6 &  0       & &  34$\pm$3 & 23$\pm$3 &  0 \\
23.0 &  52$\pm$2 & 35$\pm$2 &  2$\pm$11 & &  3$\pm$10 &  1$\pm$18 &  0       & &   4$\pm$8 &  2$\pm$11 & 0 \\
\hline
\end{tabular}
\end{minipage}
\end{table*}

\subsection{Colour Magnitude Diagrams}
\begin{figure*}
\includegraphics[width=140mm]{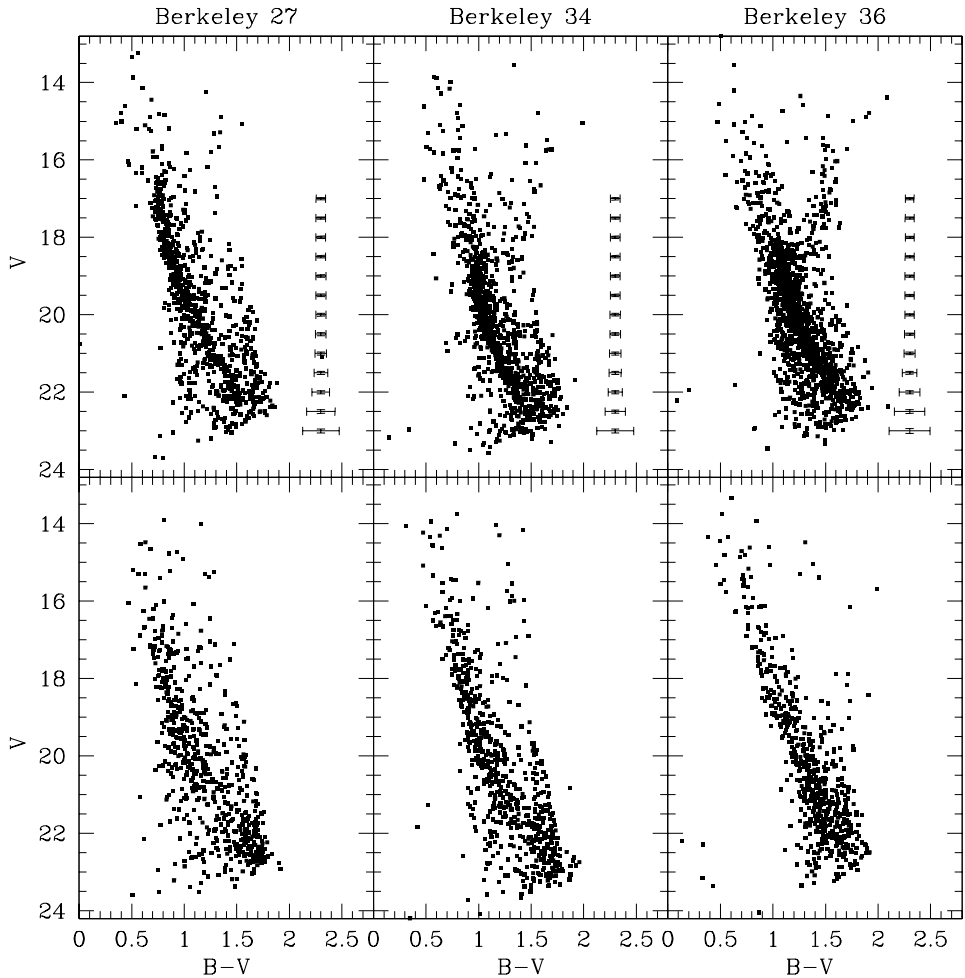}
\caption{Upper panels: CMDs of Be~27, Be~34, and Be~36 showing $V$ vs $(B-V)$. Lower panels: CMDs of the corresponding comparison field. Global photometric errors are shown on the right side of the clusters CMDs.}
\label{fig:CMDer}
\end{figure*}

\begin{figure*}
\includegraphics[width=140mm]{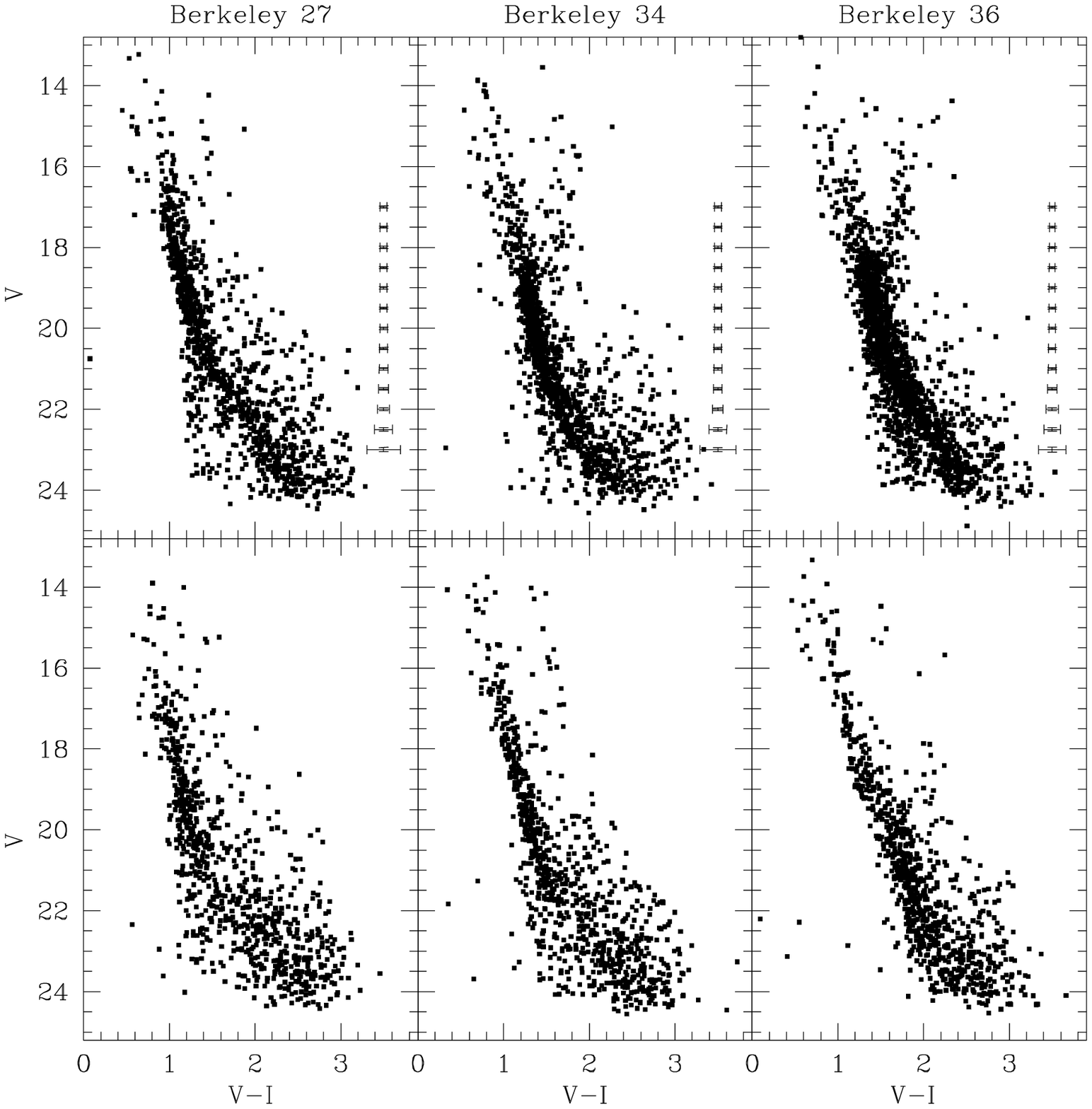}
\caption{Same as Figure~\ref{fig:CMDer} but for $V$ vs $(V-I)$.}
\label{fig:CMDer2}
\end{figure*}

The resulting CMDs for the cluster stars and the comparison field stars are shown in Figures~\ref{fig:CMDer} ($V,B-V$ plane) and \ref{fig:CMDer2} ($V,V-I$ plane). Error bars indicate the global photometric error that takes into account the instrumental error and the uncertainties on the calibration procedures. They range from about 0.03 mag at the bright limit to less than 0.1 mag around $V=24$. The three OCs main evolutionary phases are visible despite the important field contamination. In particular,  the Main Sequence (MS) is easily recognisable although its broad shape does not help in defining its features. The detailed analysis of the CMDs morphology is described in Section~\ref{sec:CMDsynth}.  

\subsection{Comparison with previous data}
\begin{figure}
\includegraphics[width=1.0\linewidth]{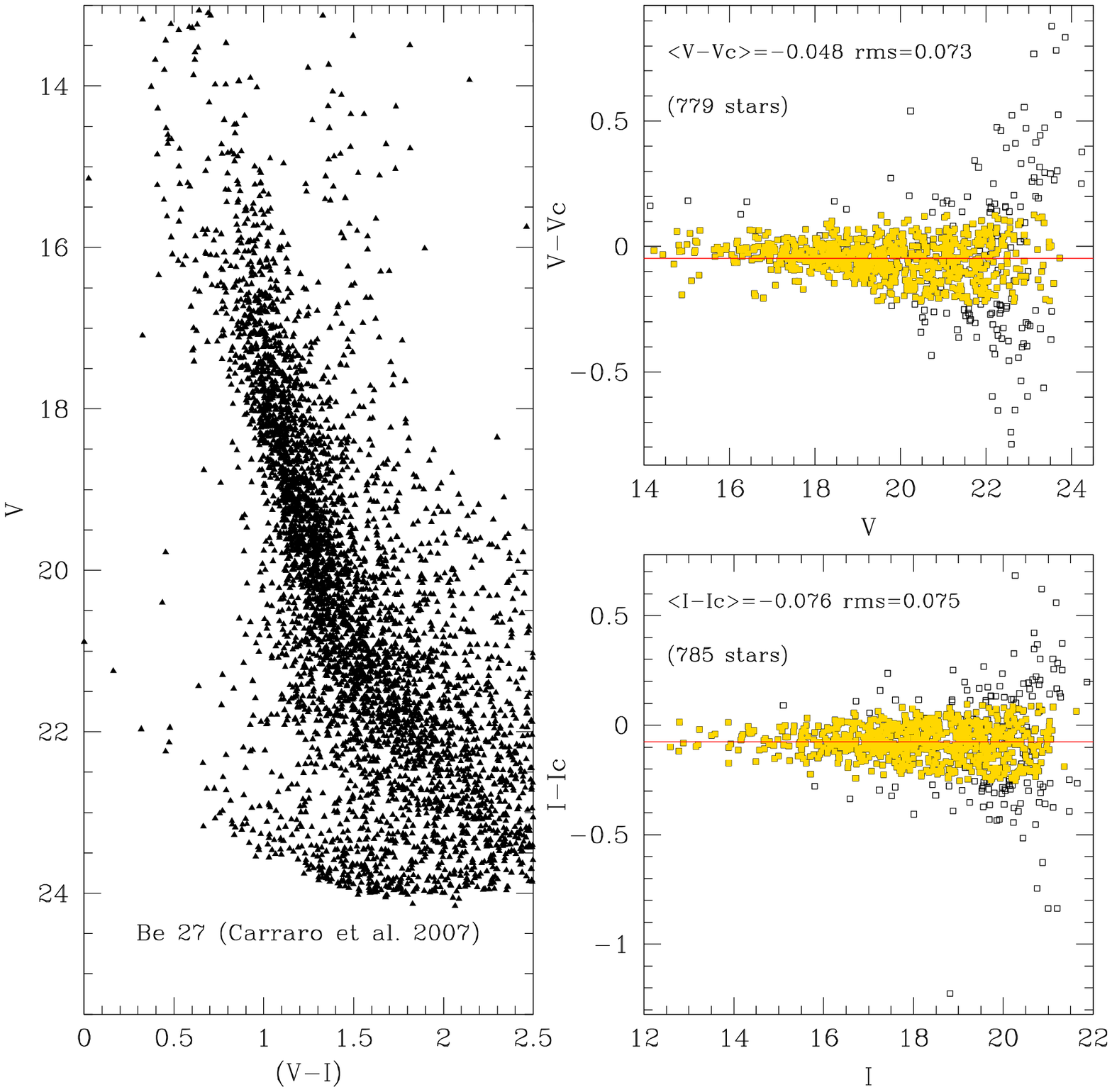}
\caption{Left panel: CMD of Be~27 by Carraro et al. (2007). Right panel: differences between our photometry and theirs in $V$ (upper panel) and $I$ (lower panel). Points are all the stars in common; filled ones are stars used to compute the mean differences (within 2$\sigma$ from the average).}
\label{fig:be27vs}
\end{figure}

\begin{figure}
\includegraphics[width=1.0\linewidth]{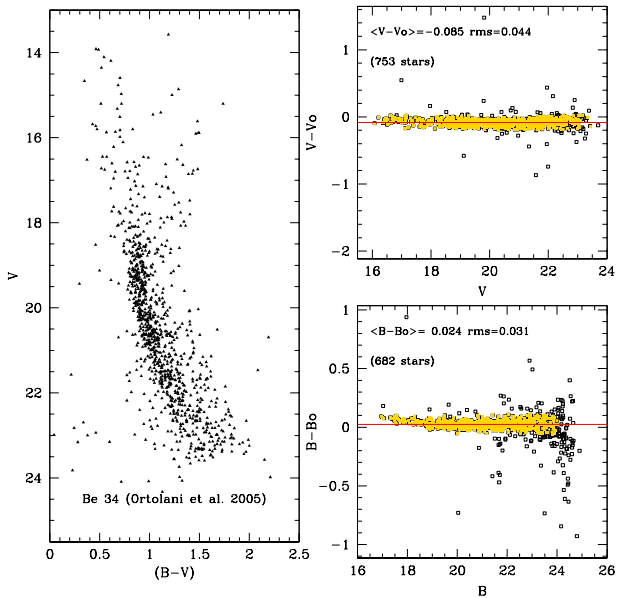}
\caption{Left panel: CMD of Be~34 by Ortolani et al. (2005). Right panel: comparison with our data for $V$ (upper panel) and $B$ (lower panel) magnitudes (see previous figure).}
\label{fig:be34vs}
\end{figure}

\begin{figure}
\includegraphics[width=1.0\linewidth]{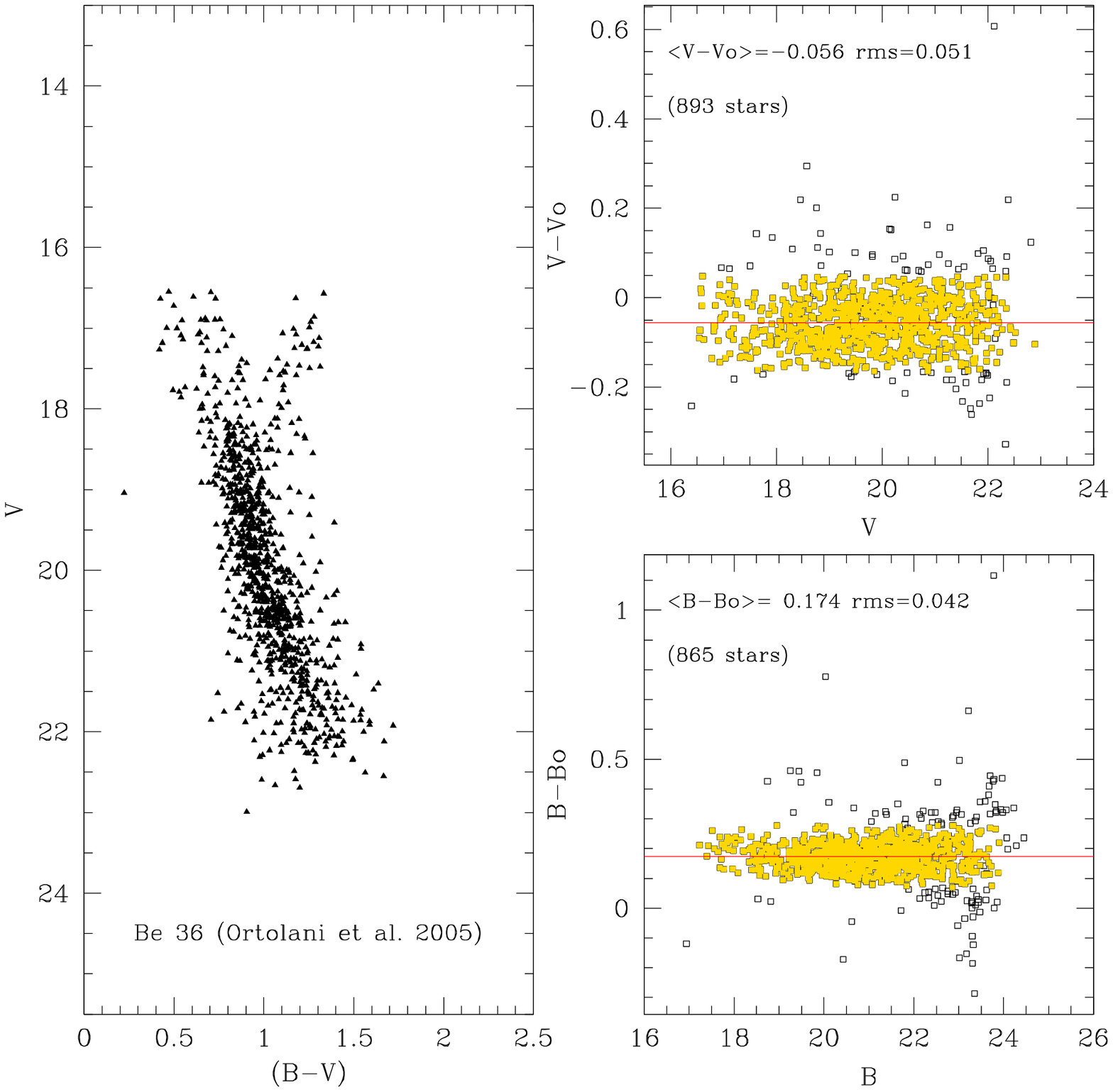}
\caption{Left panel: CMD of Be~36 by Ortolani et al. (2005) of the stars in common with our catalogue. Right panel: comparison with our data (see previous figure).}
\label{fig:be36vs}
\end{figure}

As said in Sect.~\ref{sec:intro}, Be~27, Be~34, and Be~36 were previously observed by various authors. The web database for OCs, WEBDA\footnote{http://www.univie.ac.at/webda/}, was exploited to obtain literature data. In particular Be~27 was studied by \cite{hase_04} and \cite{car_07}, Be~34 and Be~36 by \cite{hase_04} and \cite{orto_05}. The work done by \cite{hase_04} contains $B$, $V$, and $I$ photometry but they made public through WEBDA only data with $V<18$. Furthermore, for Be~27 they declare to have problems of photometric calibration.  So we decided to show a comparison only with the data from \cite{car_07} which contains $V$ and $I$ photometry for Be~27 and with the data from \cite{orto_05} that has $B$ and $V$ photometry for Be~34 and Be~36.
In Figures~\ref{fig:be27vs},~\ref{fig:be34vs},~and~\ref{fig:be36vs} the comparisons of the photometries are shown together with the literature CMDs.

Concerning Be~27 (Fig.~\ref{fig:be27vs}), the difference in the $V$ filter is, on average, $-0.04$ mag with a very shallow slope for bright stars. The difference in the $I$ filter is a bit larger, about $-0.07$ mag, leading to an average difference in  
$(V-I)$ colour of +0.03. We cannot tell if these small differences are due to our photometry or theirs since there are no other reliable data in literature for this cluster. Moreover, it is not possible to solve the issue with general considerations about our ability to calibrate the photometry to the standard system because our OCs were observed in different nights.

For Be~34 (Fig.~\ref{fig:be34vs}) and Be~36 (Fig.~\ref{fig:be36vs}) the comparison in $V$ gives an average difference of -0.085 mag and -0.056 mag respectively, while for the $B$ photometry we find a difference of +0.024 mag and +0.174 mag. For the former OC the difference in $B$ is small but becomes important in $V$, leading to a difference in $(B-V)$ of +0.11 mag. In the case of Be~36 the disagreement in $B$ is quite significant and we obtain an average difference in $(B-V)$ of +0.23 mag. The explanation of such differences is not straightforward as it is not possible to definitely distinguish if they are due to our photometry or theirs. Both our targets and theirs were observed in two different nights with no significant difference in the calibration parameters obtained for the two nights. Furthermore we had some problems in attempting the cross-correlation of our astrometrised catalogues with the pixel-coordinates ones of \cite{orto_05}. We could find a good match only dividing the whole literature catalogues in two halves: one with all the stars with x coordinates smaller than 1025 and the other one with x coordinates larger than 1025. This problem is probably due to geometrical distortion in the alignment of the two CCDs pixel coordinates performed by \cite{orto_05}. In addition we noticed that for Be~36 the original data file available through WEBDA (the same we downloaded from the Vizier portal\footnote{http://vizier.u-strasbg.fr/viz-bin/VizieR}) contains differences with respect to the CMD shown in \cite{orto_05} and with respect to the plot facilities of the WEBDA itself. Therefore we chose to show in Figure~\ref{fig:be36vs} only the CMD of the stars in common with our catalogue and we assumed that our photometry is on the standard system in the following analysis.

\section[]{Clusters centre}
\label{sec:centre}
Occasionally the cluster centre indicated in the WEBDA (coming from the \citealt{dias_02} catalogue and updates) is
offset from the true one, even by a few arcminutes, so we checked if this was the case for the three OCs. 
For each object we computed its centre as the barycentre of the stars spatial distribution on the basis of a simple statistical approach. The three clusters are all distant objects, so their apparent diameter is relatively small, giving us the chance to distinguish the central part even with the small FoV of SUSI2. From the \cite{dias_02}  catalogue  we know that the apparent diameter of the clusters (based on visual inspection) is about 2 arcmin for Be~34 and 5 arcmin for Be~36, while for Berkeley 27 the recent study by \cite{car_07} indicates a cluster radius of 3 arcmin. This means that two of our OCs are fully contained in the SUSI2 FoV, while Be~27 is slightly larger.

To identify the centre, first we restricted our analysis only to stars which belong to the upper part of the CMDs (selecting those with $V\leq22$) to have a smaller sample of objects strongly dominated by cluster stars with a relative small contamination of field stars. Then we performed a spatial selection: we computed the smallest intervals in coordinates RA and Dec that contain 70\% of stars and, with this smaller group, we iterated the computation to define a spatial region used to refine our analysis. The cluster centre is then computed as the barycentre of the final group of stars. 
\begin{table}
 \centering
 \begin{minipage}{84mm}
  \caption{J2000 RA and Dec coordinates for the three clusters. The second and third columns are the computed coordinates of the centre. The last two columns contain the previous determinations of the centre.}
  \label{tab:centre}
  \begin{tabular}{@{}cccccc@{}}
\hline
\hline
Cluster & \multicolumn{2}{c}{Centre} & & \multicolumn{2}{c}{Previous Determination\footnote{Source: WEBDA}} \\
 & RA & Dec & & RA & Dec \\
\hline
Be~27 & 06 51 21 & +05 46 07 & & 06 51 18 & +05 46 00\\
\hline
Be~34 & 07 00 23 & -00 13 56 & & 07 00 24 & -00 15 00\\
\hline
Be~36 & 07 16 24 & -13 11 35 & & 07 16 06 & -13 06 00\\
\hline
\end{tabular}
\end{minipage}
\end{table}
 
The guidelines adopted for the spatial selection are set heuristically, aiming at taking into account the clustering  level of the stars (which dominates on small scales) but also the sparse nature and asymmetric distribution of objects in OCs (which dominates on large scales). The constraints on the algorithm seemed to us a good trade-off, confirmed by the small dispersion of the results with respect to different magnitude selections (7 arcseconds for RA and 2 arcseconds for Dec).  

The results, which are the average of different selections in magnitude, are shown in Table~\ref{tab:centre}. They are slightly different from the literature ones, especially for Be~36.

\section[]{Clusters parameters using synthetic colour-magnitude diagrams}
\label{sec:CMDsynth}
The estimations of age, metallicity, distance, mean Galactic reddening, and binary fraction have been obtained comparing the observational CMDs with a library of synthetic ones, built using synthetic stellar populations \citep[see e.g., ][]{tosi_07,cig_11}. Different sets of evolutionary tracks\footnote{The Padova \citep{bre_93}, FRANEC \citep{dom_99}, and FST ones \citep{ven_98} of all available metallicities as in all the papers of the BOCCE series.} have been used to Monte Carlo generate the synthetic CMDs. The comparison between synthetic and observed CMDs is based on the CMD morphology and number counts. The best fit solution is chosen as the one that can best reproduce some age-sensitive indicators as the luminosity level of the MS reddest point (``red hook'', RH), the red clump (RC) and the Main Sequence Termination Point (MSTP, evaluated as the maximum luminosity reached after the overall contraction, OvC, and before the runaway to the red), the luminosity at the base of the red giant branch (RGB), the RGB inclination and colour, and the RC colour. The last two were used as secondary age indicators as colour properties are more affected by theoretical uncertainties, like colour transformations and the super-adiabatic convection, while luminosity constraints are more reliable. 

The most valuable age indicator is the Turn Off (TO) point, that is the bluest point after the OvC, and the RC luminosity; however, at least in the case of OCs, these phases may be very poorly populated, and identifying them is not a trivial game, especially if a strong field stars contamination is present.

In order to make a meaningful comparison, the synthetic CMDs are made taking into account the photometric error, the completeness level of the photometry, and the stellar density contrast of the open clusters population with respect to the population of the comparison field. The synthetic CMDs are combined with stars picked from an equal area of the comparison field to take the contamination into account. 

As we did in \cite{cig_11}, we first evaluated the parameters that do not depend on the evolutionary model analysis, as the binary fraction and the differential reddening. The binary fraction is estimated from the information on colour and magnitude of the cluster stars then fine-tuned, together with the differential reddening parameter, in order to match the MS width. In the analysis described in the next paragraphs the adopted differential reddening is considered as an upper limit and added as a random positive constant to the mean Galactic reddening. The luminosity of the MSTP and of the RC are effectively used to constrain the age. The estimated luminosity of the base of the RGB (BRGB), the RGB inclination and colour, and RC colour are used to select the best fit to the observational CMDs in order to estimate the mean Galactic reddening $E(B-V)$ and observed distance modulus $(m-M)_0$, and to fine tune the metallicity. The best estimate of the mean Galactic reddening is defined when the bluest upper part of the synthetic CMD MS matches the corresponding part of the observed CMD MS; the observed distance modulus is identified when the MSTP level and colour are reproduced in the synthetic CMD. We took into account the information of the complete $BVI$ photometry to constrain the metallicity \citep[see][]{tosi_07} and reduce the parameter space of our analysis: the best metallicity is defined when it is possible to reproduce at the same time the observed $B-V$ and $V-I$ CMD with the synthetic ones for appropriate distance modulus, reddening, and age. To deal with $(B-V)$ and $(V-I)$ colours we adopted the normal extinction law where $E(V-I)=1.25\times E(B-V)\times[1+0.06\times(B-V)_0+0.014\times E(B-V)]$ \citep{dean_78}.  

This procedure relies mostly on the MS fitting and the RC fitting. Hence, the main uncertainties on the results are due to the fact that the MS inclination and RC morphology and luminosity are quite sensitive to the input physics of the model and to the adopted colour transformations, and the uncertainties in defining the RC stars are not negligible for poorly populated clusters, increasing the probability of confusion with RGB and field stars biasing the age determination. In this context the ``best'' solution parameters are chosen as the ones which fit most of the visible MS shape and the assumed RC level.

We estimated the errors on the cluster parameters (mean Galactic reddening, distance modulus, and cluster age) considering the instrumental photometric error and the uncertainties of the fit analysis. The net effect of the former is an uncertainty on the luminosity level and colour of the indicators adopted. This in particular affects the mean Galactic reddening and the distance modulus estimations as they are directly defined by matching the level and colour of the upper MS and the RH and MSTP indicators of the observed CMDs with the synthetic ones. For the latter we consider the dispersion in the results arising from the fit analysis: Open Clusters offer poor statistics and important age sensitive indicators, such as the RC locus, are poorly defined, hence there is not an unambiguous solution but a range of compatible solutions. Then we select the best fitting synthetic CMD and take into account in the error budget the dispersion of the cluster parameters estimates for the different solutions. The uncertainties are taken to be of the form:
$$\sigma^2_{E(B-V)}\sim\sigma^2_{(B-V)}+\sigma^2_{fit}$$
$$\sigma^2_{(m-M)_0}\sim\sigma^2_{V}+R_V^2\sigma^2_{E(B-V)}+\sigma^2_{fit}$$
$$\sigma^2_{age}\sim\sigma^2_{fit}$$
the typical photometric error for the reddening is $\sim0.04$ and for the distance modulus $\sim0.1$ (we considered negligible the error on $R_V$); the dispersion for the fit analysis depends mainly on the uncertainty on the RC level and on the coarseness of the isochrone grid. It is of the order of $\sim0.02$ for the reddening, and ranges between 0.01 and 0.05 for the distance modulus, and about 0.2-1 Gyr for the age.

\subsection{Berkeley 27}
\begin{figure*}
\includegraphics[width=1.0\linewidth]{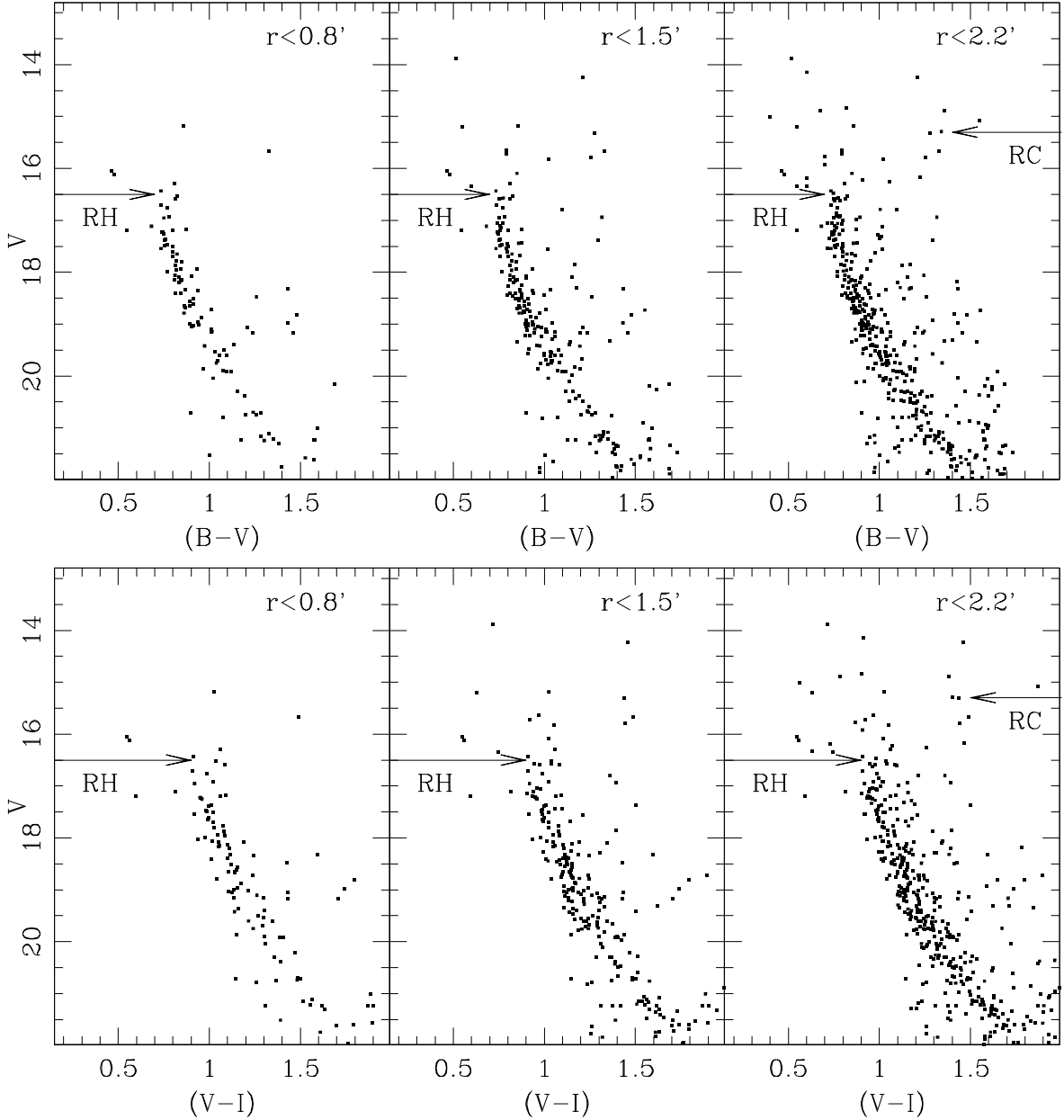}
\caption{Upper panels show the $V$ vs $(B-V)$ CMDs of Be~27 for different distances $r$ from the cluster centre. Lower panels: the same but for the $V$ vs $(V-I)$ CMDs. In the plots  the level of RH (solid arrow on the left) and RC (solid arrow on the right) are also indicated.}
\label{fig:be27rad}
\end{figure*}

\begin{figure}
\includegraphics[width=1.0\linewidth]{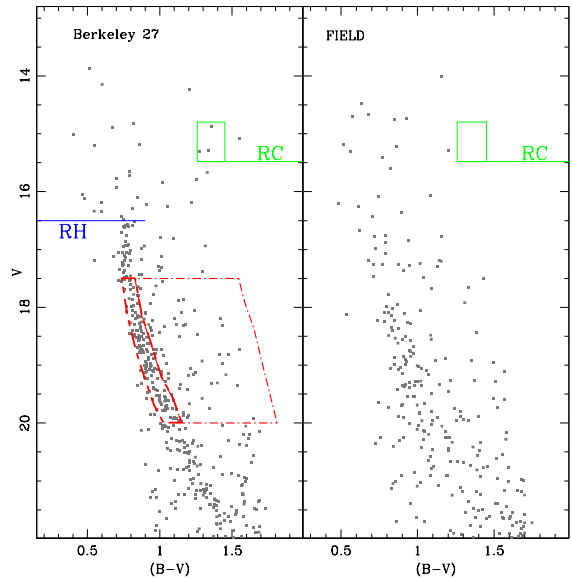}
\caption{Left panel: CMD of Be~27 for stars falling inside a region of 2.2 arcmin from the cluster centre. We indicate the luminosity level of the RH and the RC. The dotted and the dot-dashed boxes are, respectively, used to estimate the fraction of single and binary stars. Right panel: CMD of the comparison field of an equal area.}
\label{fig:be27vsfield}
\end{figure}

Be~27 is a poorly populated cluster: the contrast of member stars with the comparison field ones outlines the cluster MS but other evolutionary phases are not easily recognisable. For a more robust analysis we studied the inner part of the cluster which is less contaminated by field interlopers. Figure~\ref{fig:be27rad} shows the CMDs for different circular areas around the cluster centre: the left panels are the CMDs for the smallest area (distance from the centre $r<0.8\arcmin$) and only the MS is clearly visible. We indicate the RH level (solid arrow on the left): the MS shows a little bend toward the red just below the RH then reaches its reddest point at $V\sim16.5$. The two blue stars at $(B-V)\sim0.45$ and $V\sim16$ are probably cluster blue straggler stars (very common in OCs, see e.g., \citealt{bss} for a recent catalogue). The central panels of the same figure show the CMD for stars with a distance from the cluster centre smaller than 1.5$\arcmin$ and the right one for a distance $r<2.2\arcmin$. Concerning the RC stars there is no firm evidence from the CMDs; we define the most probable RC locus (solid arrow on the right) choosing the two stars at magnitude $V\sim15.2$ and colour $(B-V)\sim1.3-1.4$: these stars are close to the centre, hence they are more likely cluster members and have a very small difference both in $B-V$ and $V-I$ colours; in addition, our choice is in agreement with the analysis by \cite{car_07}. The RC stars are very few but still more abundant than the comparison field stars (see CMDs in Figure~\ref{fig:be27vsfield} for a circular area of $r=2.2\arcmin$). The uncertainty in defining the magnitude level of the RC directly affects the precision of the age estimate. We chose to use the CMD for $r<2.2\arcmin$ in the further analysis to limit the field contamination.

The MS appears broader than expected from photometric errors. This is probably due to two factors: one is a large fraction of binaries and the other is the presence of differential reddening. For our simulations we needed to assume a differential reddening of $\Delta E(B-V)=0.05$ mag in addition to the mean Galactic reddening.

A rough estimate of the binary fraction was obtained following the method described in \cite{cig_11}: we defined two CMD boxes, one which encloses MS stars and the other red-ward of the MS in order to cover the binary sequence (see dashed and dot-dashed lines in Figure~\ref{fig:be27vsfield}). To remove the field contamination we subtracted the contribution of field stars falling inside the same CMD boxes of an equal area of the control field. We performed the same computation on regions smaller and larger than 2.2$\arcmin$, finally ending with an estimate between 20\% and 30\%. The dispersion on the estimate is mostly due to the spatial fluctuations across the control field. Moreover these fractions are underestimated: we are missing binaries hosting  low mass star, whose properties are close to those of single stars. However, a mean fraction of 25\% appears  a reasonable trade-off and will be assumed for all the simulations.

We performed the simulations looking for the best combination of parameters keeping fixed the binary fraction and differential reddening derived above. 
The interval of confidence for the cluster age turns out between 1.2 and 1.8 Gyr. 
Concerning the metallicity, we found that all models with solar metallicity can not fit the stellar population both in $(B-V)$ and $(V-I)$ colours. Therefore we concentrated our efforts on solutions with $Z<0.02$.

The FST models with $Z=0.006$ and $Z=0.01$  (overshooting parameter $\eta=0.2$) fit reasonably well the RH and RC luminosity levels. The synthetic MS is slightly redder than the observed one in the magnitude range $18.5<V<20$ mag and this is probably due to the fact that the synthetic MS shape is too curved before the RH point. In terms of cluster parameters $Z=0.006$ implies a cluster age of $1.5\pm0.2$ Gyr, $E(B-V)=0.50\pm0.04$, and a distance modulus $(m-M)_0=13.20\pm0.13$; $Z=0.01$ implies a cluster age of $1.5\pm0.2$ Gyr, $E(B-V)=0.44\pm0.04$, and a distance modulus $(m-M)_0=13.27\pm0.13$. We can not firmly choose between the two metallicities $Z=0.006$ and $Z=0.01$: from the comparison of $(B-V)$ and $(V-I)$ CMDs with the observed ones we find a good match in both cases. This means that the metallicity estimate suffers more uncertainties, as we can not obtain a unique and independent evaluation from the $BVI$ photometry but only put an upper limit. On the other hand, the circumstance that with both metallicities we obtain the same age and distance modulus (obviously not the same reddening) emphasises the robustness of their values. 

Of the Padova models we used the ones with $Z=0.004$ and $Z=0.008$. In the first case we obtain the best match assuming a cluster age of $1.7\pm0.2$ Gyr, a reddening of $E(B-V)=0.52\pm0.04$ and a distance modulus of $(m-M)_0=13.05\pm0.13$. Both the RC and RH levels have a good fit, matching also the RC colour. As for the FST models, we find a slightly redder MS for $V>19$ mag. The difference remains also with the other tracks with $Z=0.008$. For this metallicity we estimate a cluster age of $1.7\pm0.2$ Gyr, $E(B-V)=0.44\pm0.04$, and a distance modulus of $(m-M)_0=13.10\pm0.13$. Also for the Padova models we can only put an upper limit to the cluster metallicity using $(B-V)$ and $(V-I)$ CMDs comparison, and constrain the metallicity estimate in terms of the best synthetic CMD fit. Again for age and distance we get stable solutions.

With the FRANEC models we used metallicity $Z=0.006$ and $Z=0.01$. In the former case we can match the RH and RC levels with a reasonable fit of the upper part of the MS while the lower part ($V>18.5$) has a redder $(B-V)$ colour. We determine a cluster age of $1.2\pm0.2$ Gyr, $E(B-V)=0.54\pm0.04$, and $(m-M)_0=13.1\pm0.13$. For the latter case we obtain a fit that shares the same problems of the previous one: the RH and RC levels are well matched but the lower part of the synthetic MS is redder for $V>18.5$. Accepting these differences we confirm a cluster age of $1.2\pm0.2$ Gyr with $E(B-V)=0.50\pm0.04$, and $(m-M)_0=13.10\pm0.13$. For the FRANEC models the higher metallicity ($Z=0.01$) gives a slightly better match both in $(B-V)$ and $(V-I)$, reproducing better the RH phase. As usual, the ages derived from the FRANEC models are lower than those from both the Padova and the FST ones. This is because the FRANEC tracks do not include overshooting from convective cores, while the other two sets do.

\begin{figure*}
\includegraphics[width=1.0\linewidth]{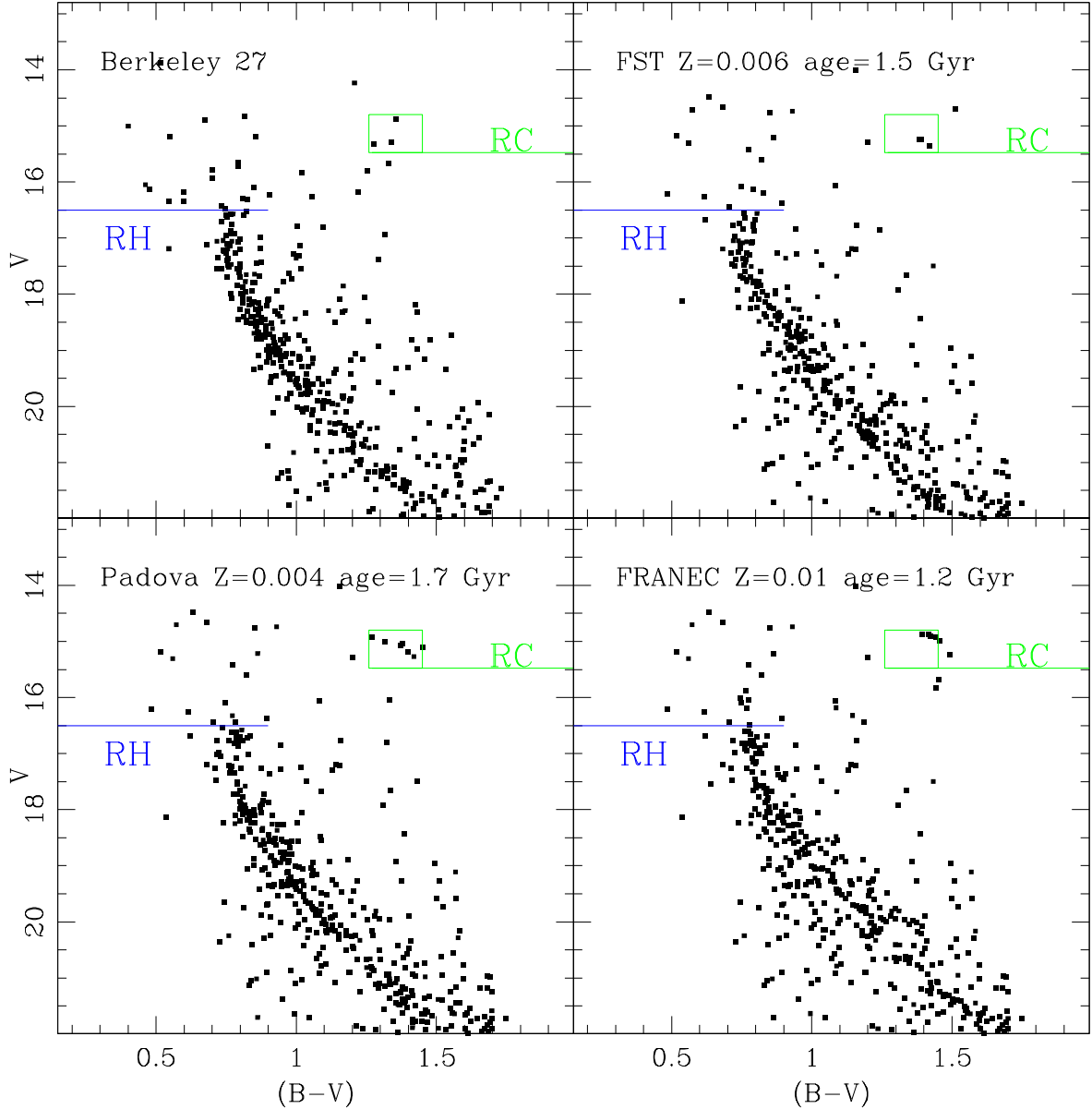}
\caption{Top left panel: CMD of stars inside 2.2$\arcmin$ radius area of Be~27. The top right panel shows the best fitting CMD obtained with FST model: $Z=0.006$, age 1.5 Gyr, $E(B-V)=0.50$, and $(m-M)_0=13.2$; the bottom left panel is the synthetic CMD obtained with Padova track: $Z=0.004$, age 1.7 Gyr, $E(B-V)=0.52$, and $(m-M)_0=13.05$; finally, the bottom right CMD has been obtained with FRANEC model: $Z=0.01$, age 1.2 Gyr, $E(B-V)=0.50$ and $(m-M)_0=13.1$.}
\label{fig:cmdsbe27}
\end{figure*}

Figure~\ref{fig:cmdsbe27} shows the comparison between the observed CMD (top left) and the best fits obtained with the three sets of tracks.

From this analysis it turns out that the FST models are the ones that best fit the observed CMD as they provide a better match of the MS shape. This restricts the age to 1.5 Gyr. Consequently the Galactic reddening is between 0.40 and 0.50 mag which nicely compares with the \cite{sch_98} estimate of 0.49 mag, while the distance modulus is between 13.2 and 13.3. 

To Be~27 \cite{car_07} assign an age of 2 Gyr, older than our estimates but still compatible with the results obtained with Padova models (the ones used by the authors). This difference is mainly due to the identification of the RC level. The cluster in fact lacks  a clear RGB and clump, leaving more uncertainties on the age determination. Restricting the comparison to the Padova models, the chosen metallicity used for the fit can explain the difference for the reddening estimate, as the photometry offset between our data and theirs is of the order of 0.03 mag for $(V-I)$: for higher metallicities the fit requires lower reddening values as the isochrone has a redder colour. We find a distance modulus larger (about 0.4 mag) and this is mainly due to the age adopted (the offset in photometry is only of the order of 0.05 mag): the higher the age the fainter the magnitude of the TO, therefore a good fit is obtained with a smaller value of the distance modulus.  

\subsection{Berkeley 34}
\begin{figure*}
\includegraphics[width=1.0\linewidth]{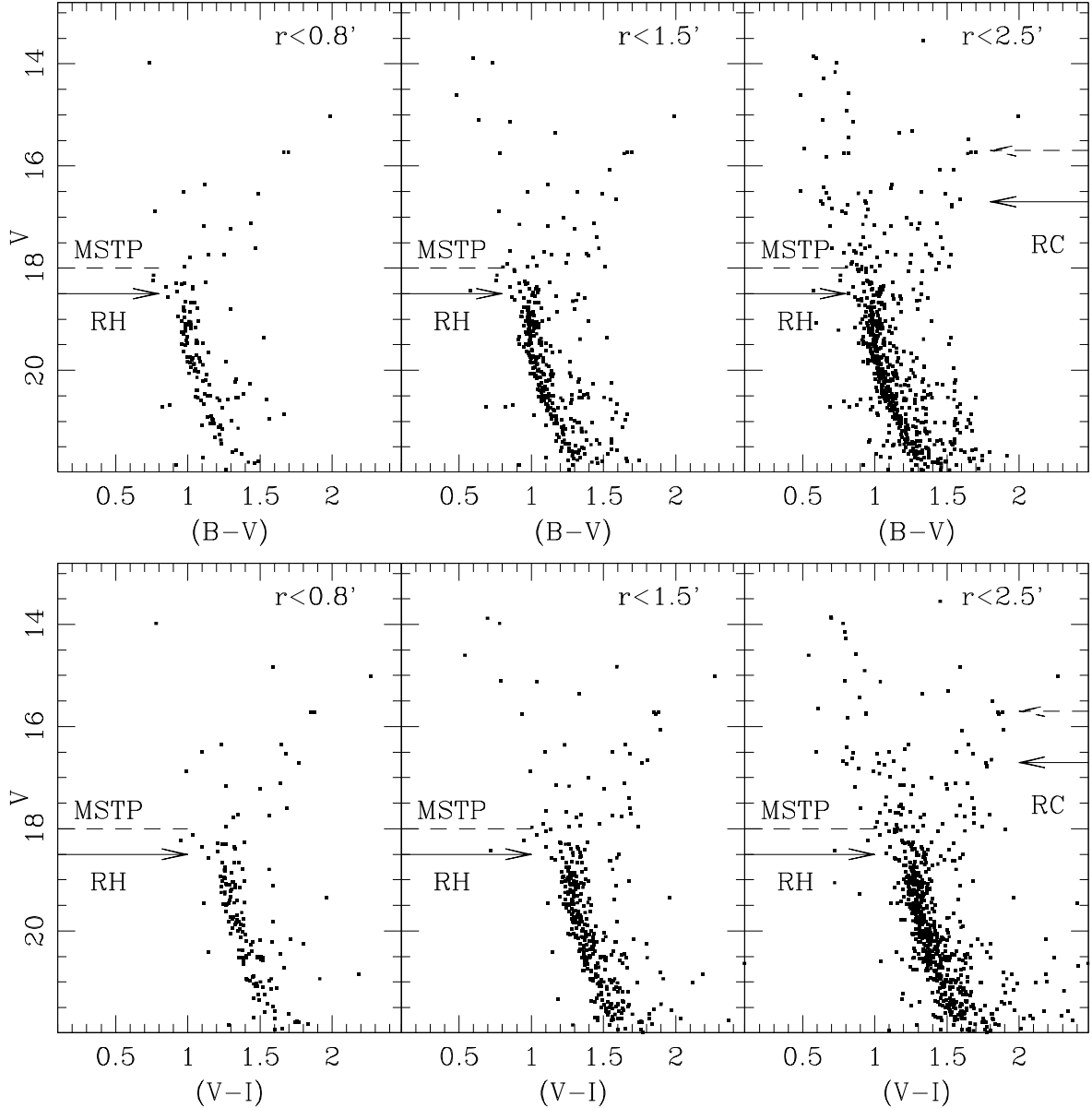}
\caption{Upper panels show the $V$ vs $(B-V)$ CMDs of Be~34 for different distances $r$ from the cluster centre. Lower panels: the same but for the $V$ vs $(V-I)$ CMDs. In the plots are also indicated the levels of RH (solid arrow on the left) and MSTP (dashed line). We indicate also the two RC levels identified: dashed (rejected) and solid (adopted) arrows on the right.}
\label{fig:be34rad}
\end{figure*}

The CMD of Be~34 is much richer than that of Be~27. In Figure~\ref{fig:be34rad} we show the $(B-V)$ and $(V-I)$ CMDs for different circular areas centred on the cluster. The plots on the left are a selection of the very central part of Be~34 (distance $r$ lower than $0.8\arcmin$): the MS is well visible and we indicate the RH level, positioned near $V$=18.5 mag and the MSTP level, set near $V\sim18.0$ mag. In the central ($r<1.5\arcmin$) and right ($r<2.5\arcmin$) panels the MS is better delineated but with a heavier contamination of field stars.
We identify two different equally probable locations for the red clump: one is the bright small group of three stars at $V\sim15.7$ mag and $(B-V)\sim1.7$ (dashed arrow on the right), the second is the fainter group (4 and more sparse stars) at $V\sim16.7$ and $(B-V)\sim1.55$ (solid arrow on the right). The uncertainty on the RC level comes from the fact that this evolutionary phase is very scarcely populated. In the first case we would estimate an older age for the cluster, as the magnitude difference between the MSTP and the RC levels is larger.  

\begin{figure}
\includegraphics[width=1.0\linewidth]{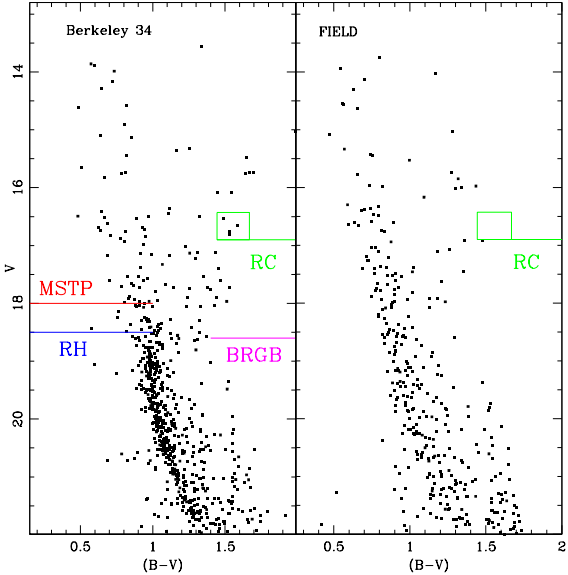}
\caption{Left panel: CMD of Be~34 for stars falling inside a region of 2.5 arcmin from the cluster centre. We indicate the luminosity level of the RH, the MSTP, the RC, and the BRGB. Right panel: CMD of the comparison field of an equal area.}
\label{fig:be34vsfield}
\end{figure}

In Figure~\ref{fig:be34rad} we show, in the left panel, the CMD of stars selected in a region within 2.5$\arcmin$ from the cluster centre (we used this selection for the following analysis) and in the right panel the comparison field of an equal area. We indicate also the RH, MSTP, and RC magnitude levels. The RGB is difficult to recognise. It is populated by a little bunch of stars that runs red-ward of $(B-V)$=1.5 and brighter than $V$=18.0. We identify the base of the RGB (BRGB) at level $V$=18.6 (see Figure \ref{fig:be34vsfield}). In the comparison field there is no star with $(B-V)>1.5$ and no counterpart at the RC levels defined above.

As for Be~27, the MS appears broader than expected from the photometric errors: presumably differential reddening and binaries play a non negligible role in shaping the MS appearance. For our simulations we took into account a differential reddening of at least 0.05 mag. The percentage of binaries was computed using the same approach done for Be~27,  finding an average fraction of 27\%.

In order to put limits on the cluster age and metallicity, the CMD of the region within 2.5$\arcmin$ is compared with our synthetic CMDs. We found that models with metallicity $Z<0.02$ are in agreement with both $(B-V)$ and $(V-I)$ therefore we  discarded models with solar metallicity. 

If we adopt the brighter RC level estimation we find that the synthetic CMDs can match well the indicators levels (RH, MSTP and RC) but with a worse fit for the lower MS ($V>18.0$) and for the RGB and RC colours (too blue). Even if the colour indicators are prone to greater uncertainties, as explained at the beginning of this section, in our opinion these discrepancies come from an incorrect age estimation: as the age of the stellar population increases the colour extension of the sub giant branch (SGB) becomes shorter. We thus took into account also this age sensitive indicator, looking for a reasonable match of the distance in colour between the MS and the BRGB. In addition these solutions cannot fit the very bright ($V\sim15$) and red ($B-V\sim2.0$) star, that seems to be an RGB cluster member. Our final choice is therefore to identify the RC at $V\sim17.2$ and $(B-V)\sim1.55$.

For the FST models (with overshooting parameter $\eta=0.2$) we find a reasonable agreement between synthetic and observed CMDs for a cluster age of $2.1\pm0.2$ Gyr with both metallicities $Z=0.006$ and $Z=0.01$. The RH, MSTP, BRGB, and RC levels are well matched with a proper fit of the MS and of the RGB shapes. The better match is obtained with the model with $Z=0.01$: the bright red member mentioned above suggests an RGB inclination which better matches the metal-rich model. The chosen binary fraction seems in agreement with the observations: the broad lower part of the MS is well reproduced. The reddening and distance modulus assigned for the model with $Z=0.006$ are $E(B-V)=0.62\pm0.04$ and $(m-M)_0=14.2\pm0.13$. For $Z=0.01$ we estimated $E(B-V)=0.57\pm0.04$ and $(m-M)_0=14.3\pm0.13$. From this analysis we find that the models with $Z=0.006$ and $Z=0.01$ provide good matches both in the $(B-V)$ and in the $(V-I)$ CMDs, leaving open the choice between these two metallicities. 

Using the Padova tracks with $Z=0.004$ and $Z=0.008$ we obtain in both cases a good match for RH, MSTP, and BRGB magnitudes and colours, as well as a reasonable fit for the MS shape and RGB colour and inclination. Also in this case the best match is obtained using the metal-richer model. For $Z=0.004$ we infer a cluster age of $2.5\pm0.2$ Gyr, $E(B-V)=0.64\pm0.04$, and a distance modulus $(m-M)_0=14.1\pm0.13$. With $Z=0.008$ we derive a cluster age of $2.3\pm0.2$ Gyr, $E(B-V)=0.59\pm0.04$, and a distance modulus $(m-M)_0=14.2\pm0.13$. Also for the Padova models with sub-solar metallicity we obtain a good match in the $(B-V)$ and in the $(V-I)$ CMDs, hence we can not firmly choose between $Z=0.004$ and $Z=0.008$.   

With the FRANEC models we obtain a younger age estimate for the cluster. These models in fact do not consider overshooting and this naturally leads to a lower age prediction. The younger age required to fit the luminosity constraints results in a synthetic CMD that has a too red RGB and a too faint BRGB. For $Z=0.006$ we estimate a cluster age of $1.6\pm0.2$ Gyr, $E(B-V)=0.67\pm0.04$, and a distance modulus $(m-M)_0=14.2\pm0.13$. With $Z=0.01$ we obtain a cluster age of $1.6\pm0.2$ Gyr, $E(B-V)=0.65\pm0.04$, and a distance modulus $(m-M)_0=14.2\pm0.13$. In this case the higher metallicity ($Z=0.01$) gives a slightly better match both in $(B-V)$ and $(V-I)$, with a better fit of the upper MS morphology.  

\begin{figure*}
\includegraphics[width=1.0\linewidth]{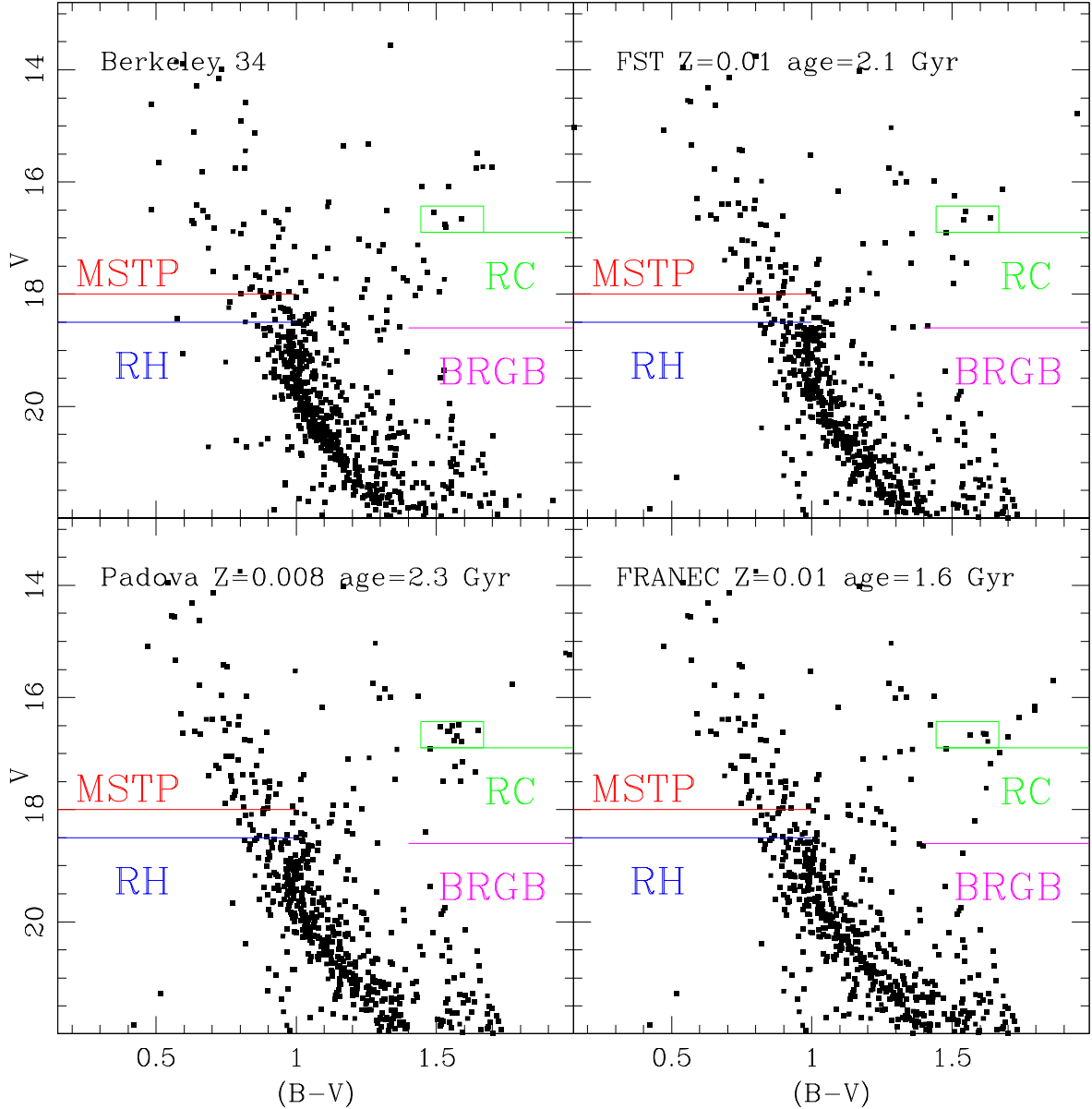}
\caption{Top left panel: CMD of stars inside 2.5$\arcmin$ radius area of Be~34. The top right panel shows the best fitting CMD obtained with FST model: $Z=0.01$, age 2.1 Gyr,  $E(B-V)=0.57$ and $(m-M)_0=14.3$; the bottom left panel is the synthetic CMD obtained with Padova track: $Z=0.008$, age 2.3 Gyr, $E(B-V)=0.59$, and $(m-M)_0=14.2$; finally, the bottom right CMD has been obtained with FRANEC model: $Z=0.01$, age 1.6 Gyr, $E(B-V)=0.65$ and $(m-M)_0=14.2$.}
\label{fig:cmdsbe34}
\end{figure*}

Figure~\ref{fig:cmdsbe34} shows the best fitting CMD for each set of tracks and the corresponding parameters. We prefer the FST models as they give a better description of the CMD morphology as a whole. We find in fact that the Padova models predict a MS shape too curved before the RH point. The FRANEC models, instead, give a good match of the magnitude indicators  but a worse fit for the MS shape and RGB inclination. With this assumption the age of Be~34 is estimated as 2.1 Gyr, with a range in reddening between 0.57 and 0.62 \citep[similar to the][value of 0.68]{sch_98}\footnote{
Unfortunately, given the very low latitude of all three OCs, these reddening values cannot be trusted to give the real asymptotic reddening, and cannot give a firm constraint as in more favourable cases.} and a distance modulus between 14.2 and 14.3.

\cite{orto_05} assign to this cluster an age of 2.3 Gyr, which is in agreement with our estimation. In particular it coincides with the one we obtained with the Padova models (the ones used by them). However, their choice of RC level does not seem to agree with either one of our two possibilities. A non negligible difference is found in the reddening and distance modulus determination. In the first case the discrepancy can be explained in terms of differences between our photometries (see Sect. 2.4). For the distance modulus the differences in the photometries can not explain such discrepancy: we notice that they chose a MSTP level about half magnitude brighter with respect to our analysis, hence they determined a smaller distance modulus.

\subsection{Berkeley 36}
\begin{figure*}
\includegraphics[width=1.0\linewidth]{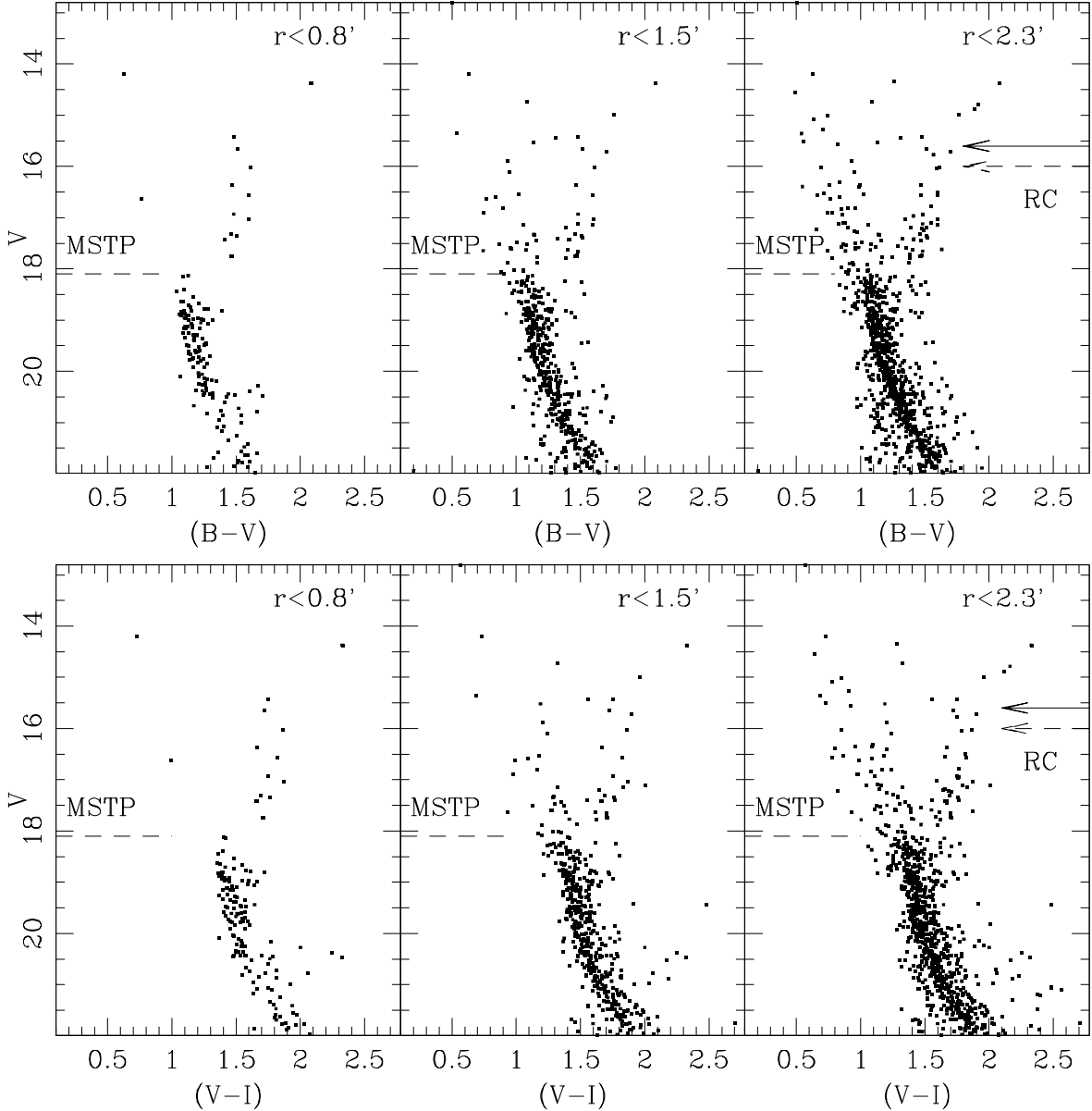}
\caption{Upper panels show the $V$ vs $(B-V)$ CMDs of Be~36 for different distances $r$ from the cluster centre. Lower panels: the same but for the $V$ vs $(V-I)$ CMDs. In the plots we indicate the levels of MSTP (dashed line) and RC. The dashed arrow is for the rejected level (see text) and the solid arrow is for the adopted RC level.}
\label{fig:be36rad}
\end{figure*}

\begin{figure}
\includegraphics[width=1.0\linewidth]{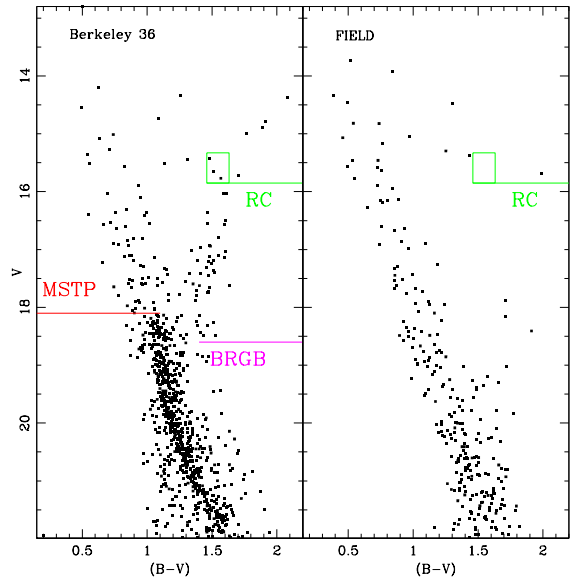}
\caption{Left panel: CMD of Be~36 for stars falling inside a region of 2.3 arcmin from the cluster centre. We indicate the luminosity level of the MSTP, BRGB, and the RC. Right panel: CMD of an equal area of the comparison field.}
\label{fig:be36vsfield}
\end{figure}

Be~36 is the richest cluster of the group. The CMDs in Figure~\ref{fig:be36rad} clearly show the MS, the MSTP ($V\sim18.1$), and the RGB for different distances from the cluster centre. The contamination from field stars is evident particularly in the central and right panel: the MS is blurred and the region above the MSTP is dominated by field interlopers (together with the cluster blue straggler stars, very common in OCs, see e.g., \citealt{bss} for a recent catalogue). Also for this cluster we restricted our analysis to a small area of 2.3$\arcmin$ of radius to maximise the membership likelihood. Even within this restricted area we can still notice an important field contamination but without losing the evidence of the CMD features: the MSTP at $V\sim18.1$ and the BRGB at the magnitude level of  $V\sim18.6$ (see Figure~\ref{fig:be36vsfield}). We also notice a small gap at $V\sim18.7$ which could be associated to an RH phase; however, further investigations discarded this hypothesis. The RGB is quite evident, running red-ward of $(B-V)=1.5$ and reaching $V\sim14.5$ with a very red member at $B-V\sim2.2$. The field contamination along the RGB seems very modest, from comparison to an equal area of the external field (Figure~\ref{fig:be36vsfield}). Yet, the RC level is not so evident: we adopted as RC the small group of stars (two) located at $V\sim16.0$ (in Figure~\ref{fig:be36rad} we indicate with the dashed arrow the probable RC). However, even though we obtained a good fit of the RC and MSTP levels and of the MS shape, we could not obtain a good description of the RGB phase, too red in the synthetic CMDs. This is due to a too extended SGB phase, suggesting that we are adopting a too young age for the cluster. To help choose the best solution ever without a firm evidence of RC stars, we compared the CMD of Be~36 with those of two of the oldest clusters inside the BOCCE project: Berkeley 17 (Be~17) and Berkeley 32 (Be~32). Be~17 is among the oldest OCs of the Galaxy, with an age in the range 8.5-9.0 Gyr \citep{bra_06} while Be~32 is 5-5.5 Gyr old \citep{tosi_07}. They both have sub-solar metallicity, as expected for Be~36 from previous analysis. 

\begin{figure*}
\includegraphics[width=1.0\linewidth]{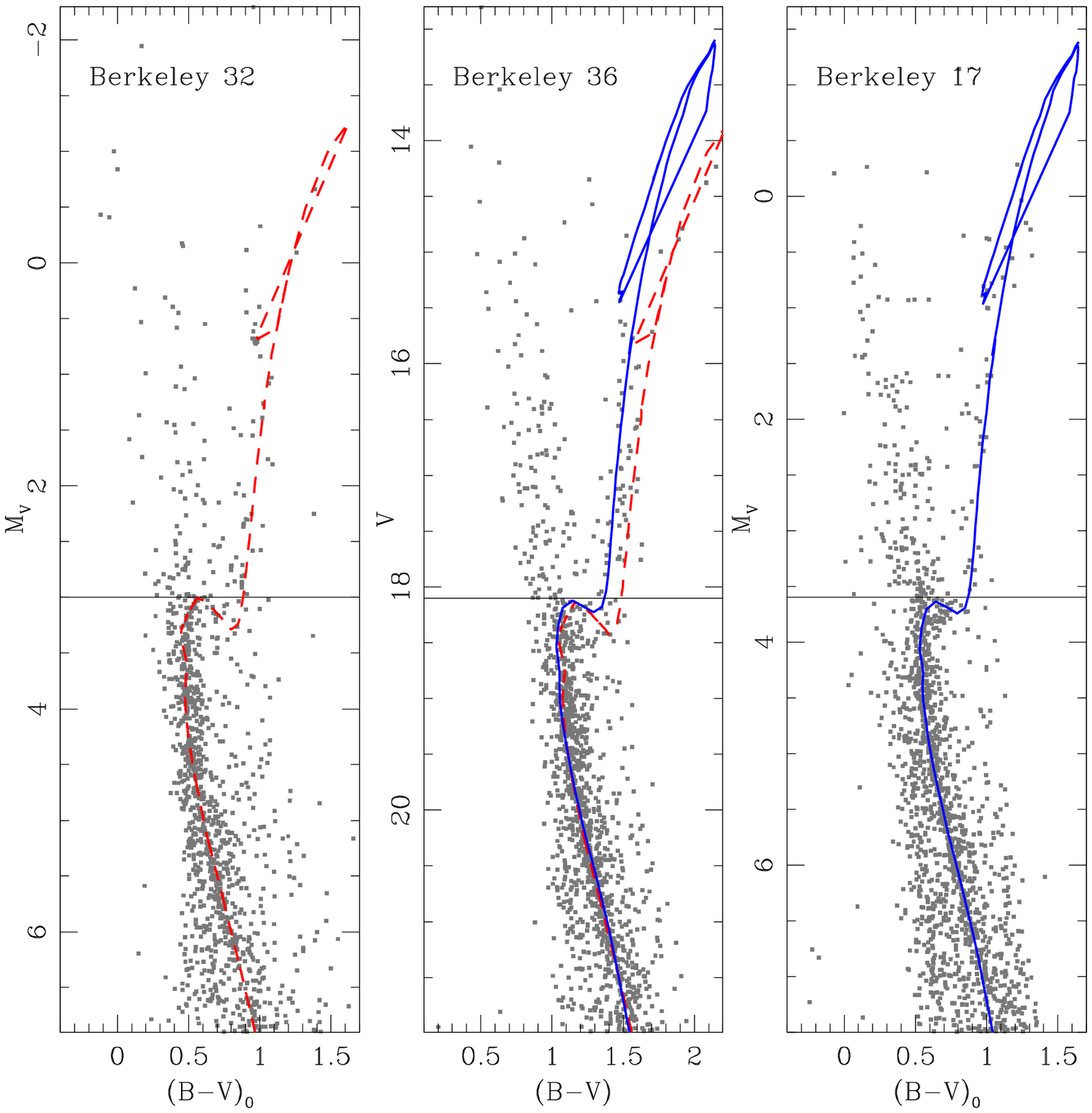}
\caption{Left panel: CMD $M_V$,$(B-V)_0$ of Be~32. The dashed line is the best fit solution described in Tosi et al. 2007: $Z=0.008$, age 5.2 Gyr, $E(B-V)=0.12$, and $(m-M)_0=12.6$ for the Padova models. Central panel: CMD of Be~36 with the overplot of the best fit isochrones of Be~32 (dashed line) and Be~17 (solid line). Right panel: CMD $M_V$,$(B-V)_0$ of Be 17. The solid line is the best fit solution for the Padova models described in Bragaglia et al. 2006: $Z=0.008$, age 8.5 Gyr, $E(B-V)=0.62$, and $(m-M)_0=12.2$. We used different limits on the magnitude (y-axis) for the three plots in order to visually align the evolutionary MSTP phase of the clusters but preserving the magnitude and colour range for a easier comparison. The solid horizontal lines set the MSTP level.}
\label{fig:vsbe32be17}
\end{figure*}

In Figure~\ref{fig:vsbe32be17} we show a comparison of the CMDs of Be~32, Be~36, and Be~17. In the left and right panels we present the CMDs of Be~32 and Be~17 using absolute magnitude $M_V$ and intrinsic colour $(B-V)_0$. We used different limits on the magnitude (y-axis) to visually align the luminosity level of the evolutionary MSTP phase of the clusters, preserving the magnitude and colour range in order to properly compare the CMDs. We also show the isochrones which best fit the clusters according to our analysis (dashed line for Be~32 and solid line for Be~17). We overplot them on the CMD of Be~36 after a proper alignment in colour and magnitude. While both isochrones fit well the upper and lower MS, they bracket the RGB of Be~36 on the red and blue side.

This indicates that Be~36 is in an evolutionary status intermediate between that of Be~32 and Be~17. In particular, we can discard a cluster age younger than about 5 Gyr, as it would imply a more extended SGB and a redder RGB, while the older isochrone shown in the comparison sets a upper limit (8.5 Gyr) to the cluster age. Assuming the ages of Be~32 and Be~17 as limits for Be~36, the corresponding most probable RC locus for this cluster is at $V\sim15.5$ and $(B-V)\sim1.55$ (solid arrow on the right in Figure~\ref{fig:be36rad}). The three stars enclosed in the box in Figure~\ref{fig:be36vsfield} are very few but likely to be cluster members as they are positioned near the cluster centre. 

Having so decided the RC position and an age range, we applied the usual method of analysis, taking into account the very scattered characteristic of the CMD. We adopted a higher differential reddening of 0.15, the only viable solution to reproduce the MS spread, and a binary fraction of 25\%. Given the more scattered appearance, these values have larger uncertainties than for the two other clusters. Keeping fixed these parameters we investigated the possibility to fit simultaneously the MSTP, BRGB, and RC luminosities by adjusting the age, the mean Galactic reddening, and the distance modulus. 

\begin{figure}
\includegraphics[width=1.0\linewidth,height=2.5\linewidth]{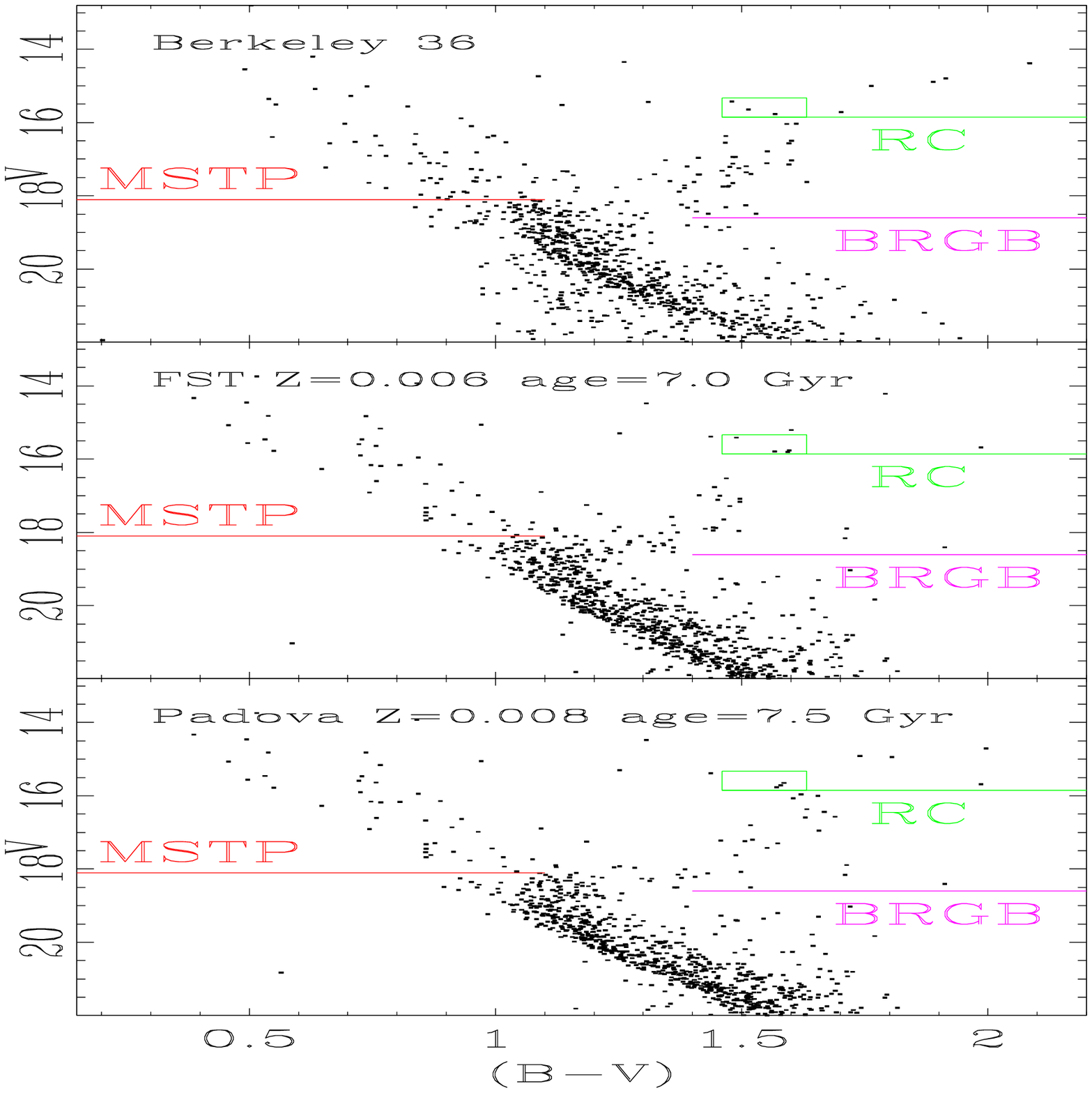}
\caption{Top panel: CMD of stars inside 2.3$\arcmin$ radius area of Be~36. The central panel shows the best fitting CMD obtained with the FST models: $Z=0.006$, age 7.0 Gyr,  $E(B-V)=0.53$ and $(m-M)_0=13.15$; the bottom panel is the synthetic CMD obtained with the Padova tracks: $Z=0.008$, age 7.5 Gyr, $E(B-V)=0.51$, and $(m-M)_0=13.1$.}
\label{fig:cmdsbe36}
\end{figure}

Also for Be~36 we restricted our analysis to models with sub-solar metallicity: a metallicity of $Z=0.02$ can not match at the same time $B-V$ and $V-I$, predicting a RGB with a strong inclination in the upper part. In contrast with what was found for Be~27 and Be~34, all the explored models predict a lower MS slightly bluer than observed. We could not use the FRANEC models as they have incomplete evolutionary tracks for ages older than 5 Gyr for subsolar metallicities. Figure~\ref{fig:cmdsbe36} displays the best fitting CMD for each set of tracks compared with the observational CMD (upper panel).

The FST models can reproduce quite well the magnitudes and colours of the indicators, even though they predict a bluer MS for $V>21$. We find a better match for models with $Z=0.006$. For $Z=0.01$, when the synthetic $B-V$ CMD is correct, the $V-I$ always turns out to be slightly bluer than observed. The best solution is obtained for the models with $Z=0.006$, a cluster age of $7.0\pm1.0$ Gyr, a mean reddening of $E(B-V)=0.53\pm0.04$, and a distance modulus $(m-M)_0=13.15\pm0.13$. For $Z=0.01$ we find the same age of $7.0\pm1.0$ Gyr, $E(B-V)=0.48\pm0.04$, and  $(m-M)_0=13.19\pm0.08$. 

Using the Padova models we find a good match for the MSTP, BRGB and RC levels with a better description of the MS (bluer only for $V>21.5$). The best matches are obtained with models with $Z=0.008$, when the synthetic CMDs can match the observed one both in $B-V$ an $V-I$ at the same time. We find a cluster age of $7.5\pm1.0$ Gyr, $E(B-V)=0.51\pm0.04$, and $(m-M)_0=13.1\pm0.13$; the synthetic CMD obtained with these parameters reproduces quite well the MS and RGB shape and colour, even if it can not reproduce correctly the over-density observed at $V\sim17$ along the RGB.

The comparison with previous results \citep{orto_05} shows a significant discrepancy in the cluster age and therefore in the determination of cluster reddening and observed distance modulus. This is in part due to the choice of the MSTP level and in part to the disagreement in the photometries (see Sect. 2.4). Concerning the age they set the MSTP level half a magnitude brighter than our estimation, adopting the same RC level we use for the analysis. This implies a younger age and a smaller distance modulus estimations. The difference in the reddening estimates is mainly due to the remarkable disagreement in the photometries.

\section{Conclusions}
\label{sec:sum}
The purpose of this paper is to add additional empirical  information to the models of the Galactic disc structure and chemical evolution. We studied three distant open clusters toward the anti-centre direction using SUSI2@NTT $BVI$ photometry. With these data we obtained CMDs one magnitude deeper with respect to the ones found in literature. This aspect is especially relevant for the more distant and reddened clusters Be~34 and Be~36, for which we could obtain more precise data for the lower MS. The analysis was carried on using the synthetic CMDs technique that allowed us to infer a confidence interval for age, metallicity, binary fraction, reddening, and distance for each clusters. We used three different sets of stellar tracks (Padova, FST, FRANEC) to describe the evolutionary status of the clusters in order to minimise the model dependence of our analysis. We found that:
\begin{itemize}
\item Be~27 is located at about 4.0-4.5 kpc from the Sun (assuming the normal extinction law $R_V=A_V/E(B-V)=3.2$). Its position in the Galactic disc is at $R_{GC}\sim11.8-12.2$ kpc and 185-205 pc above the plane (assuming $R_\odot=8$ kpc as in our previous works). The resulting age varies between 1.2 and 1.7 Gyr, depending on the adopted stellar model, with better fits for ages between 1.5 and 1.7 Gyr. A metallicity lower than solar seems preferable. The mean Galactic reddening $E(B-V)$ is between 0.44 and 0.54 and we estimate a (lower limit) fraction of binaries of about 25\%.
\item Be~34 is 6-7 kpc away from the Sun, with a distance from the Galactic centre of about 14.0-14.6 kpc and located 220-240 pc above the plane. The age is between 1.5 and 2.5 Gyr, with better fits in the age range 2.1-2.5 Gyr. The metallicity for this cluster is lower than solar; the mean Galactic reddening $E(B-V)$ is between 0.57 and 0.64. The estimated binary fraction for this cluster is about 27\%.
\item Be~36 is about 4.2 kpc away from the Sun. Its distance from the Galactic centre is $R_{GC}\sim11.3$ kpc and it lies 40 pc below the plane. This cluster shows a broad differential reddening up to $+0.15$, adding uncertainty to the interpretation of the cluster parameters. The best fitting age is between 7.0 and 7.5 Gyr with a preference for models with a metallicity lower than solar and higher than $Z=0.004$. The reddening estimate is  $E(B-V)\sim0.5$, while the binary fraction is of the order of 25\%.
\end{itemize}
Poorly populated clusters such as Be~27 have a very loose and barely observable RC and RGB, condition that adds uncertainties on the study of the evolutionary status of the objects. On the other hand, clusters like Be~34 and Be~36 have a much more evident RGB but suffer from a greater contamination of field stars and a stronger differential reddening: in this case the RC determination is strongly affected by these two aspects. Relaxing the assumptions on the RC position could change noticeably the cluster age for Be~27, for which we can only rely on the MSTP and MS shape, while for Be~34 and Be~36, the additional information on the  well populated RGB better constrains the analysis.
A robust determination of the three clusters parameters would require additional information on cluster membership for evolved and MSTO stars. This is obtainable in the immediate future measuring radial velocities of at least many tens of stars,
or we can wait for the results of the Gaia astrometric satellite, with precise individual distances and proper motions.

For all the three clusters we found a metallicity lower than solar, even if we were not able to unambiguously tell   if Z=0.004, 0.006, 0.008, or 0.01 (depending on the track used) is to be preferred. This conforms to their Galactocentric distance. Only high resolution spectroscopy  of these clusters  will be able to definitely determine the metallicity value. Given the relatively faint magnitudes even of the red giants, an 8-10m telescope will be necessary; it is however an important piece of information for the chemical modelling of the Galaxy.

\section*{Acknowledgements}
We thank Paolo Montegriffo, whose software for catalogue matching we consistently use for our work.
We are grateful to the referee, Bruce Twarog, for his encouraging and always constructive comments.
For this paper we used  the VizieR catalogue access tool (CDS, Strasbourg, France), WEBDA, and NASA's Astrophysics Data System.




\bsp

\label{lastpage}

\end{document}